\documentclass[prl,superscriptaddress,twocolumn,longbibliography]{revtex4-2}

\pdfoutput=1
\usepackage{dcolumn,array}
\usepackage{bm}
\usepackage{subcaption}
\usepackage{amsmath,amsfonts,amssymb,graphics,graphicx,epsfig,color,times,natbib}
\usepackage{tikz}
\usepackage{pgfplots}
\usepackage{verbatim}
\usepackage{url}
\usepackage[unicode=true,
 bookmarks=true,bookmarksnumbered=false,bookmarksopen=false,
 breaklinks=false,pdfborder={0 0 1},backref=false,colorlinks=true]
{hyperref}
\hypersetup{
 linkcolor=magenta, urlcolor=blue, citecolor=blue, pdfstartview={FitH}, hyperfootnotes=false, unicode=true}

\usepackage{algorithm}
\usepackage{algpseudocode}
\usetikzlibrary{shapes.geometric, arrows}
\usepackage{ragged2e}

\newcommand{\e}{\mathrm{e}}
\newcommand{\eps}{\varepsilon}

\newcommand{\Norm}[2][]{||{#2}||_{#1}}

\newcommand{\E}{\mathbb{E}}

\renewcommand{\Pr}{\mathrm{Pr}}
\newcommand{\mc}[1]{\mathcal{#1}}
\newcommand{\mr}[1]{\mathrm{#1}}

\newcommand{\ket}[2][]{{|{#2}\rangle_{#1}}}

\usepackage{mathtools}
\usepackage{amsmath}
\usepackage{amsthm}
\usepackage{times}

\newtheorem{theorem}{Theorem}

\newtheorem{corollary}{Corollary}

\begin{document}

\title{Classical Verification of Quantum Learning Advantages with Noises}

\author{Yinghao Ma}
 \affiliation{Center for Quantum Information, IIIS, Tsinghua University, Beijing 100084, China}
 
\author{Jiaxi Su}
 \affiliation{Center for Quantum Information, IIIS, Tsinghua University, Beijing 100084, China}
 
\author{Dong-Ling Deng}
\email{dldeng@tsinghua.edu.cn}
\affiliation{Center for Quantum Information, IIIS, Tsinghua University, Beijing 100084, China}
\affiliation{Hefei National Laboratory, Hefei, China}
\affiliation{Shanghai Qi Zhi Institute, Shanghai 200232, China}

\begin{abstract}
Classical verification of quantum learning allows classical clients to reliably leverage quantum computing advantages by interacting with untrusted quantum servers. 
Yet, current quantum devices available in practice suffers from a variety of noises and whether existed classical verification protocols carry over to noisy scenarios remains unclear. 
Here, we propose an efficient classical error-rectification algorithm to reconstruct the noise-free results given by the quantum Fourier sampling circuit with practical constant-level noises. In particular, we prove that the error-rectification algorithm can restore the heavy Fourier coefficients by using a small number of noisy samples that scales logarithmically with the problem size.
We apply this algorithm to the agnostic parity learning task with uniform input marginal and prove that this task  can be accomplished in an efficient way on noisy quantum devices with our algorithm. In addition, we prove that a classical client with access to the random example oracle can verify the agnostic parity learning results from the noisy quantum prover in an efficient way, under the condition that the Fourier coefficients are sparse. Our results demonstrate the feasibility of classical verification of quantum learning advantages with noises, which provide a valuable guide for both theoretical studies and practical applications with current noisy intermediate–scale quantum devices.
\end{abstract}
% \date{\today}
\maketitle

Quantum machine learning \cite{Biamonte2017Quantum,DasSarma2019Machine,Dunjko2018Machine,Cerezo2022Challenges} studies the interplay between quantum computing and machine learning, promising unprecedented potential to solve problems that are otherwise unattainable for classical computers. 
The core idea is to leverage quantum resources—such as entanglement, nonlocality, and contextuality—to enhance, speed up or innovate machine learning. There has been a number of notable works \cite{Lloyd2014Quantum,Rebentrost2014Quantum,Gao2018Quantum,Liu2021Rigorous} utilizing quantum algorithms \cite{Grover1996fast,Brassard2002Quantum,Harrow2009Quantum} to accelerate specific classical machine learning problems, showing potential exponential quantum advantages. 
In addition, previous studies have also extended classical learning theories, such as probably approximately correct (PAC) learning \cite{Valiant1984Theory} and agnostic learning \cite{Haussler1992Decision,Kearns1994Efficient}, to the quantum domain 
\cite{Bshouty1998Learning,Atici2007Quantum,Servedio2004Equivalences,Montanaro2012quantum,Arunachalam2021Two,Caro2024Classical,Arunachalam2018Optimal,Atici2005Improved,Zhang2010improved,Cross2015Quantum,Grilo2019Learning}.
These efforts aim to establish rigorous bounds on the time and sample complexities of quantum learning algorithms, investigating the potential and limitations of quantum models. 
Among the various approaches within quantum  learning, quantum Fourier sampling (QFS) \cite{Bernstein1993Quantum} has emerged as a powerful tool, especially in the context of learning algorithms that involve boolean functions with a sparse Fourier representation.

Despite its theoretical promise, the practical implementation of quantum learning algorithms faces noteworthy challenges, particularly due to the presence of noises in current quantum hardware. At present, quantum devices are noisy, and  different noises can severely degrade the performance of quantum algorithms. This limitation underscores the importance of developing robust algorithms that can function effectively even in the presence of noises, a crucial step for the near-future deployment of quantum  learning in real-world applications. Quantum error correction \cite{Shor1995Scheme,Steane1996Simple,Knill1998Resilient,Terhal2015Quantum} may address general noise issues, but it  requires that the noise strength is below a certain stringent threshold and a huge number of additional qubits, rendering its applications in practical quantum learning scenarios unfeasible with current noisy intermediate–scale quantum devices \cite{Preskill2018Quantum}.

In the near future, quantum computers are most likely noisy, expensive and mainly available on the cloud \cite{2024IBM, Wurtz2023Aquila, 2024QMware, 2024Quafu}. As a result, reliable schemes that enable classical clients to delegate computation and learning tasks to untrusted quantum servers and verify the corresponding results are highly desirable. There have been several works demonstrating how clients with classical computers can verify the computing results from a quantum server \cite{Gheorghiu2018Verification,Mahadev2018Classical,Fitzsimons2017Private,Broadbent2009Universal}. For the verification of learning tasks, recently a visionary work by Caro \textit{et al.} \cite{Caro2024Classical} has extended the framework for interactive proof systems from classical computational theory \cite{Goldwasser2019knowledge, Shamir1992IP} and machine learning \cite{Goldwasser2021Interactive} to quantum machine learning.
In particular, this work considered problems such as agnostic learning of parities and Fourier-sparse functions, demonstrating that a classical verifier could efficiently obtain a sufficiently good solution by verifying the correctness of the Fourier coefficients provided by the noise-free quantum prover. However, this idealized assumption does not hold in practical settings, where noises are inevitable at the current stage.

\begin{figure*}[]
    \includegraphics[width=\textwidth]{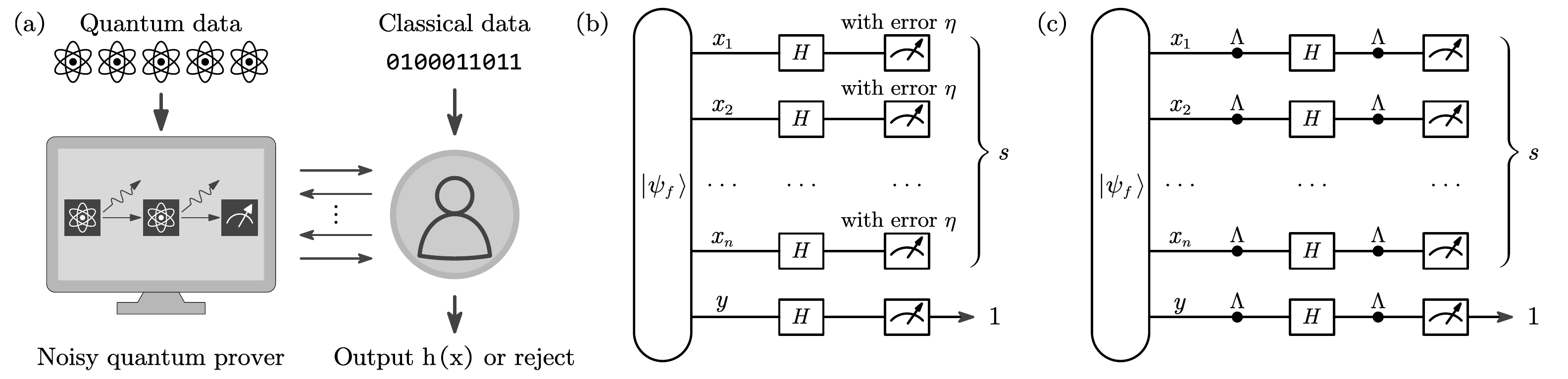}
    \caption{
    \justifying
    (a) A sketch of classical verification of quantum learning with noises. We focus our discussion on the scenario where a classical verifier interacts with an untrusted noisy quantum prover, with the verifier and prover accessing classical and quantum data, respectively. The goal of the verifier is to complete the learning task through this interaction. Based on the outcomes of the interaction, the verifier can either output the learned result $h(x)$ or reject the prover.
    (b) Quantum Fourier sampling (QFS) circuit with measurement noise, where a random bit flip with probability $\eta$ is applied independently on the measurement result for each of the first $n$ qubits.
    (c) QFS circuit with depolarization  noise, where each black node denotes  a depolarization channel  $\Lambda_{\mr{dep}}(\rho)=(1-\eta_{\mr{dep}})\rho+\eta_{\mr{dep}}I/2$, where $\eta_\text{dep}$ denotes the depolarization strength and $I$ is the identity two-by-two matrix \cite{Nielsen2010Quantum}. }
    \label{Fig.circuit}
\end{figure*}

In this paper, we focus on a more practical scenario where the quantum prover may only possess a noisy quantum computer [Fig. \ref{Fig.circuit} (a)], and propose an efficient classical error-rectification algorithm to reconstruct the corresponding noise-free results from noisy data given by the quantum server. In particular, we consider the the quantum Fourier sampling problem and prove that the error-rectification algorithm can restore the heavy Fourier coefficients by using a small number of noisy samples that scales logarithmically with the problem size. We apply this algorithm to the agnostic parity learning task  with uniform input marginal that is delegated to a untrusted noisy quantum prover. We prove that: (i) this task can be accomplished in an efficient way on noisy quantum devices
with the introduced error-rectification algorithm; (ii)  a classical client with access to the random example oracle can
verify the agnostic parity learning results in an efficient way, under the condition that the Fourier coefficients are sparse. We introduce a concrete interactive proof protocol and prove its completeness and soundness.

\textit{Notations and framework.}---
We start with the concept of (proper) agnostic PAC learning \cite{Haussler1992Decision,Kearns1994Efficient}, where the learner is asked to find an almost optimal approximation function within a given hypothesis class. Given a distribution $\mathcal{D}_x$ over an input space $\mathcal{X}_n = \{0,1\}^n$ and a target function $f: \mathcal{X}_n \to \{0,1\}$, the loss function for a hypothesis function $h: \mathcal{X}_n \to \{0,1\}$ is defined as $L_f(h)=\Pr_{x\sim\mathcal{D}_x}[h(x) \neq f(x)]$. For the $\alpha$-agnostic learning, the goal is to find a hypothesis $h_0$ in a specific hypothesis class $\mathcal{H}$ such that $L_f(h_0)\le \alpha\cdot \min_{h\in \mathcal{H}}L_f(h)+\eps$, where $\eps$ is some constant parameter. In this paper, we focus on the task of $1$-agnostic ($\alpha=1$) parity learning under a uniform input distribution $\mc U_n$. The parity function $\chi_s$ is defined as $\chi_s(x) = (-1)^{s\cdot x}$, where $s\cdot x=\sum_{i=1}^n s_i x_i \mod 2$, and the hypothesis class is defined as $\mathcal{H}=\{s\cdot x\mod 2:s\in\mc X_n\}$. 
Parity learning is closely related to the Fourier coefficients of boolean functions, defined as $\hat{g}(s) = \frac{1}{2^n}\sum_{x\in\mathcal{X}_n} g(x)\chi_s(x)$ for $g(x)=1-2f(x)$. By learning the Fourier coefficients of $g(x)$, we obtain the term with the largest Fourier coefficient, which is associated with the parity that best approximates the function $g(x)$. As a result, the task of parity learning effectively reduces to learning the Fourier coefficients of the target function.

To learn the Fourier coefficients, we utilize different types of oracles that provide access to the function $f$. The membership query oracle returns the value of $f(x)$ when queried with a specific input $x$ \cite{Valiant1984Theory,Angluin1988Queries,Goldreich1989Hardcore,Kushilevitz1991Learning}. Whereas, the random example oracle returns a pair $(x, f(x))$, where $x$ is sampled uniformly at random from $\mathcal{X}_n$ \cite{Valiant1984Theory}. For classical learning, previous works \cite{Goldreich1989Hardcore,Kushilevitz1991Learning} have demonstrated that efficient classical algorithms exist for learning $\hat{g}(s)$ using the membership query oracle by recursively identifying the heavy Fourier coefficients. However, no efficient classical algorithms have been discovered for the random example oracle so far. The quantum analogue of the random example oracle, known as the quantum example oracle \cite{Bshouty1998Learning}, returns a quantum superposition of the form
\begin{equation}
    \ket{\psi_f}=\frac{1}{2^{n/2}}\sum_{x\in\mc X_n}\ket{x,f(x)}
\end{equation}
when queried. Notice that a measurement in the computational basis of $\ket{\psi_f}$ reduces the quantum example oracle to the random example oracle, therefore the former is at least as powerful as the later.

The QFS algorithm allows quantum prover with access to the quantum example oracle to efficiently learn the Fourier coefficients of the target function, making itself a central subroutine in many quantum learning algorithm, including quantum parity learning. The noise-free circuit of QFS operates as following \cite{Bernstein1993Quantum}: given a quantum example $\ket{\psi_f}$ as input, we apply Hadamard gate to each of the $n+1$ qubits and measure them in the computational basis. The output state after the Hadamard gates reads:
\begin{equation}
    H^{\otimes (n+1)}\ket{\psi_f}=\frac{1}{\sqrt{2}}(\ket{0,0}+\sum_{s\in\mc X_n}\hat{g}(s)\ket{s,1}).
\end{equation}
The measurement result of the last qubit gives $y=1$ with probability $\frac{1}{2}$. Conditioned on $y=1$, the probability distribution of the first $n$ qubit is $p_0(s)=|\hat g(s)|^2$, whereas when  $y=0$ all the first $n$ qubits collapse to state $|0\rangle$.
Therefore by using QFS, we can sample from probability distribution $p_0$ with constant probability. Each execution of the QFS circuit requires one copy of the quantum example $\ket{\psi_f}$ and $O(n)$ single-qubit gates.

\textit{QFS with noises.}--- 
We first extend the QFS algorithm to the noisy scenario. We  consider the independent bit flip noise model, where a random bit flip with probability $\eta$ is applied independently on the measurement result for each of the first $n$ qubits [Fig. \ref{Fig.circuit} (b)]. For this circuit we still have the measurement result of the last qubit giving $y=1$ with probability $\frac{1}{2}$, but due to noises, the probability distribution of the first $n$ qubits conditioned on $y=1$ becomes:
\begin{equation}
    p_{\eta}(s)=\sum_{s'\in\mc X_n}\eta^{d_H(s,s')}(1-\eta)^{n-d_H(s,s')}p_0(s'),
\end{equation}
where $d_H(s,s')=\Norm[1]{s-s'}$ denotes the Hamming distance between $s$ and $s'$. Therefore, the problem reduces to how to reconstruct the original probability distribution $p_0$ from samples drawn from the noisy probability distribution $p_{\eta}$. We introduce a classical error-rectification algorithm to tackle this problem and prove the following theorem (throughout this paper, the notation $\tilde O$ hides log factors for $k$).

\begin{theorem}
    Let $(\theta,\delta)\in(0,1)^2$. For $\eta\le \frac{1}{10}\theta$, there exists an efficient classical algorithm that takes $k=O\left(\frac{\log(n/\delta)}{\theta^2}\right)$ noisy samples from $p_{\eta}(s)$ and finds all possibly large components in $p_0(s)$ with success probability at least $1-\delta$. In particular, such an algorithm produces a set $L$ with $|L|\le\lfloor\frac{2}{\theta}\rfloor$ satisfying $H=\{s\in\mc X_n|p_0(s)\ge \theta\}\subseteq L$ with success probability no less than $1-\delta$. 
    The algorithm's time and space complexities are $\tilde O\left(\frac{n^2\log(n/\delta)}{\theta^3}\right)$ and $O\left(\frac{n\log(n/\delta)}{\theta^2}\right)$, respectively.
    \label{T noise1}
\end{theorem}

\begin{proof}
    The rigorous proof of this theorem is lengthy and technical, and we thus leave it to Supplementary Materials Sec. II. We introduce a concrete algorithm that fulfill the task. Our basic idea in designing this algorithm is to maintain a list $L$ of all possibly heavy Fourier coefficients, and search recursively. At the $m$-th iteration, the algorithm focuses on the first $m$ qubits and the probability distribution of their measurement outcomes. We denote $p_m(s_m)$ to be the original probability distribution for the first $m$ bits, and let $H_m=\{s_m\in\{0,1\}^m|p_m(s_m)\ge \theta\}$ to be the heavy Fourier coefficients. Our goal is to find $L_m$ satisfying $H_m\subseteq L_m$. 

    Our algorithm runs as the following (see the pseudocode for this algorithm in the Supplementary Materials Sec. II). Initially, let $L_0=H_0=\{e\}$ with $e$ denoting an empty string of $\{0,1\}^*$. At step $m$, assuming we have $H_{m-1}\subseteq L_{m-1}$ and $|L_{m-1}|\le\lfloor\frac{2}{\theta}\rfloor$, and let $L'_m=L_{m-1}\times\{0,1\}$, namely $L'_m$ contains all concatenations of terms in $L_{m-1}$ and $0$ or $1$, hence $H_m\subseteq L'_m$. In the next step, we estimate the probability of each term in $L'_m$, and select about half of them to be in $L_m$. For each noisy sample $t$ drawn randomly from $p_{\eta}$, we match it with the closest (in Hamming distance) element in $L'_m$ with respect to its first $m$ bits. If there are multiple elements that are equally close, the algorithm selects one at random. Then, we compute the empirical distribution of the matching result, and selects the first $\lfloor\frac{2}{\theta}\rfloor$ elements with the largest empirical distribution to be in $L_m$. The algorithm outputs $L=L_n$ at the end.

    To show that $H_m\subseteq L_m$, we consider the first $m$ bits $t_m$ of a noisy sample $t\in T$ with the corresponding original noiseless sample being $s_m\in L'_m$. The probability for mismatching it with another $s_m'\in L_m'$ can be upper bounded by $\eta$ since the noise must flip at least half of the different bits between $s_m$ and $s'_m$ for a mismatch (see Lemma S3 in Supplementary Materials Sec. II). Thus, the probability of a correct match is at least $1-|L'_m|\eta\ge\frac{3}{5}$. As a result, for $s_m\in H_m$ with $p_m(s_m)\ge\theta$, with high probability its empirical distribution will be no less than $\frac{1}{2}\theta$, and $s_m$ will be contained in $L_m$. This leads to the conclusion that $H\subseteq L$. After some technique calculations as shown in the Supplementary Materials Sec. II, we arrive at the conclusion that the success probability $\Pr(H\subseteq L)\ge 1-2n\e^{-k\theta^2/200}$, and $k=O\left(\frac{\log(n/\delta)}{\theta^2}\right)$ noisy samples suffices.
\end{proof}

Theorem \ref{T noise1} provides an efficient classical error-rectification algorithm for QFS circuit with independent  noises to identify the heavy Fourier coefficients of the original probability distribution $p_0(s)=|\hat g(s)|^2$, by using a small number of samples from the noisy probability distribution $p_{\eta}(s)$. 
Notably, this method preserves the original QFS circuit design and does not increase the required number of qubits or gates, making it applicable to NISQ devices that suffer from system sizes and noises.  Under the condition that the noise strength does not exceed a constant level, the algorithm can restore the output results with a precision that matches the level of the noise.
We note that this error-rectification algorithm is not limited to QFS and generally applicable to a variety of scenarios involving noises, regardless of whether the task is quantum or classical. It is of independent interest.
In addition, Theorem \ref{T noise1} requires that bit flip errors occur randomly and independently for each qubit. This condition can be relaxed, as long as the error strength is smaller than a constant threshold value (see Supplementary Materials Sec. II). 

The identified heavy Fourier coefficients, together with a small number of noiseless classical random examples, enable us to estimate $p_0(s)$ for $s\in L$ using definition $\tilde{p}_0(s)=\E_x[g(x)\chi_s(x)]$ and generate an estimation $\tilde{p}_0$ of $p_0$ as shown in the following Corollary $\ref{T noise2}$:

\begin{corollary}
    Let $f:\mc X_n\to\{0,1\}$ be a boolean function and $(\eps,\delta)\in(0,1)^2$. There exists an efficient classical algorithm that takes $k=O\left(\frac{\log(n/\delta)}{\eps^4}\right)$ samples from noisy QFS circuit with error strength $\eta\le\frac{1}{10}\eps^2$ and $k'=O\left(\frac{\log(1/\eps\delta)}{\eps^2}\right)$ noiseless classical random examples to find a succinctly represented estimation $\tilde{g}$ of $\hat g$ such that $\Norm[0]{\tilde{g}}\le\frac{2}{\eps^2}$ and $\Norm[\infty]{\hat g-\tilde{g}}\le \eps$ with success probability at least $1-\delta$. The algorithm's time and space complexities are $\tilde O\left(\frac{n^2\log(n/\delta)}{\eps^6}\right)$ and $O\left(\frac{n\log(n/\delta)}{\eps^4}\right)$, respectively.
    \label{T noise2}
\end{corollary}

\begin{proof}
    We sketch the major steps here and leave the technical details to Supplemental Material Sec. III. Use the algorithm in Theorem \ref{T noise1} (with $\theta$ replaced  by $\eps^2$) to find $L$ satisfying $H=\{s\in\mc X_n|p_0(s)\ge \eps^2\}\subseteq L$ with success probability $1-\frac{\delta}{2}$. Let $T'$ be the list of classical random examples, define:
    \begin{equation}
        \tilde{g}(s)=\left\{\begin{array}{cl}
             \displaystyle \frac{1}{k'}\sum_{x\in T'}g(x)\chi_s(x), & s\in L\\
             \displaystyle 0, & s\notin L.
        \end{array}\right.
    \end{equation}
    By definition $\Norm[0]{\tilde{g}}\le|L'|\le \frac{2}{\eps^2}$. For $s\notin L$, $|\hat g(s)-\tilde{g}(s)|=|\hat g(s)|\le\eps$. For $s\in L$, by the Hoeffding inequality: $\Pr(|\hat g(s)-\tilde{g}(s)|>\eps)\le2\e^{-\frac{1}{2}k'\eps^2}$.
    Let $2\e^{-\frac{1}{2}k'\eps^2}\le\frac{\eps^2\delta}{4}$, so $k'=O\left(\frac{\log(1/\eps\delta)}{\eps^2}\right)$, then the success probability obeys $\Pr(\Norm[\infty]{\hat g-\tilde{g}}\le \eps)\ge 1-\frac{\delta}{2}-|L|\cdot\frac{\eps^2\delta}{4}\ge 1-\delta$.
\end{proof}

The above results can be readily extended to the depolarizing noise model for the QFS circuit as shown in Fig. \ref{Fig.circuit} (c). The effect of each depolarizing channel can be regarded as a probabilistic combination of the original state with probability $1-\eta_{\mr{dep}}$ and maximal mixed state $I/2$ with probability $1-\eta_{\mr{dep}}$. For each qubit, the final measurement gives the original result with probability $(1-\eta_{\mr{dep}})^2$, or a random $0/1$ result with probability $1-(1-\eta_{\mr{dep}})^2$. It is equivalent to a flip error with probability $\eta_{\mr{eff}}=\frac{1}{2}(1-(1-\eta_{\mr{dep}})^2)=\eta_{\mr{dep}}-\frac{1}{2}\eta_{\mr{dep}}^2$ on the measurement result. Unlike the previous simplified noise model, the last qubit ($y$ qubit) now also has flip error. Thus, conditioned on the case $y=1$, we find that the probability distribution of the first $n$ qubit reads:
\begin{gather}
    p_{\mr{dep}}(s)=\sum_{s'\in\mc X_n}\eta_{\mr{eff}}^{d_H(s,s')}(1-\eta_{\mr{eff}})^{n-d_H(s,s')}p_{0,\mr{eff}}(s'),\\
    \eta_{\mr{eff}}=\eta_{\mr{dep}}-\frac{1}{2}\eta_{\mr{dep}}^2,p_{0,\mr{eff}}=(1-\eta_{\mr{eff}})p_0(s)+\eta_{\mr{eff}}\delta_{s,0}.
\end{gather}
Noting that this is exactly equivalent to the simplified noise model if we replace $\eta$ and $p_0$ by $\eta_{\mr{eff}}$ and $p_{0,\mr{eff}}$, respectively. Hence, we can use the same methods in Theorem \ref{T noise1} and Corollary \ref{T noise2} to do error-rectification. More generally, any single-qubit error can be equivalently considered as a measurement outcome flip error, similar to the depolarizing noise model shown above, thereby allowing for the same methods for error-rectification.

\textit{Agnostic parity learning with noises.}---We move on to show how the error-rectification algorithm for QFS circuit with noises can be applied to solve the agnostic parity learning problem. As mentioned above, the learner is asked to find an optimal hypothesis $h(x)=s\cdot x$ to approximate the target function $f(x)$. We demonstrate that it is possible to achieve $1$-agnostic parity learning using a noisy quantum device with access to the quantum example oracle together with a classical computer with access to the random example oracle. Our results are summarized in the following theorem.

\begin{theorem}
    \label{T pac}
    Let $f:\mc X_n\to\{0,1\}$ be a boolean function and $(\eps,\delta)\in(0,1)^2$. There exists an efficient classical algorithm that takes $k=O\left(\frac{\log(n/\delta)}{\eps^4}\right)$ samples from noisy QFS circuit with error strength $\eta\le\frac{1}{10}\eps^2$ and $k'=O\left(\frac{\log(1/\eps\delta)}{\eps^2}\right)$ classical random examples to output $s_0\in\mc X_n$, such that $L_f(s_0\cdot x)\le\min_{s\in\mc X_n}L_f(s\cdot x)+\eps$ with success probability at least $1-\delta$. 
    The algorithm's time and space complexities are $\tilde O\left(\frac{n^2\log(n/\delta)}{\eps^6}\right)$ and $O\left(\frac{n\log(n/\delta)}{\eps^4}\right)$, respectively.
\end{theorem}

\begin{proof} 
    We outline the main idea here (see Supplementary Materials Sec. IV for more details). We exploit the algorithm introduced in Corollary \ref{T noise2}, from which we obtain an estimation $\tilde g$ of $\hat g$ with success probability at least $1-\delta$ such that $\Norm[\infty]{\hat g-\tilde{g}}\le \eps$. Let $s_0\in\arg\max_s\tilde{g}(s)$ and $s_1\in\arg\max_s\hat{g}(s)$, then $\hat g(s_0)\ge\tilde{g}(s_0)-\eps\ge\tilde{g}(s_1)-\eps\ge\hat{g}(s_1)-2\eps=\max_s\hat g(s)-2\eps$.
    Notice that the loss function can be written as: $L_f(s\cdot x)=\Pr_{x}(s\cdot x\ne f(x))=\Pr_{x}(\chi_s(x)\ne g(x))=\frac{1}{2^n}\sum_{x\in\mc X_n}\frac{1-g(x)\chi_s(x)}{2}=\frac{1-\hat{g}(s)}{2}$.
    Therefore we have $L_f(s_0\cdot x)\le\min_{s\in\mc X_n}L_f(s\cdot x)+\eps$.
\end{proof}

We note that 1-agnostic parity learning is at least as hard as learning parity with noise, which is widely believed to be classically intractable from random examples \cite{Lyubashevsky2005Parity,Regev2009Lattices}. 
Theorem \ref{T pac} guarantees that agnostic parity learning can be achieved in a efficient way  on noisy quantum devices with the help of classical computers. In addition, the sample, time, and space complexities scale at most polynomially with the system size, for both classical and quantum computers. The adaption of classical error-rectification algorithm into quantum agnostic learning is advantageous, especially for applications of near term noisy intermediate-scale quantum devices. From another perspective, our introduced error-rectification algorithm would make certain quantum agnostic learning approaches noisy resilient. In the following, we propose a protocol on how to verify quantum learning with noises based on the above discussion.

\textit{Classical verification of quantum learning with noises.}---
We now discuss on how a classical client verify quantum learning delegated to an untrusted noisy quantum server.
We consider a scenario where a classical verifier with access to the random example oracle interacts with a noisy quantum prover provided the quantum example oracle. We focus on the case where the target function $f$ is promised to have no non-zero Fourier coefficients smaller than $\tau$, namely for arbitrary $s\in\mc X_n$, ether $\hat g(s)=0$ or $|\hat g(s)|\ge\tau$. Based on the learning algorithm described in Theorem \ref{T pac},  we introduce in the following theorem a classical-quantum interactive proof protocol for $1$-agnostic parity learning. Our results are summarized in Theorem \ref{Thm-IntProof}.

\begin{theorem}\label{Thm-IntProof}
    Let $(\eps,\delta)\in(0,1)^2$ and $f:\mc X_n\to\{0,1\}$ be a boolean function that has no non-zero Fourier coefficients smaller than $\tau$. There exists an interactive proof system $(V,P)$ that can achieve agnostic parity learning using only one round of communication, where $V$ denotes the classical verifier with access to the random example oracle and $P$ represents the noisy quantum prover with access to the quantum example oracle. Such a proof system is both complete and sound: 
    \begin{itemize}
        \item \textbf{Completeness:} If the noise strength of  $P$ is below $O(\eps^2)$, the probability of $V$ obtaining a correct answer after interacting with $P$ is at least $1-\delta$.
        \item \textbf{Soundness:} For any prover $P'$, even with unlimited computational power and full information about the target function $f$, the probability of $V$ obtaining an incorrect answer after interacting with $P'$ is at most $\delta$.
    \end{itemize}
    Here for the agnostic parity learning the correct answer refers to a $s_0\in\mc X_n$ such that $L_f(s_0\cdot x)\le\min_{s\in\mc X_n}L_f(s\cdot x)+\eps$.
\end{theorem}

\begin{proof}
    We give the main idea here and provide more details in Supplementary Materials Sec. V. We introduce a concrete interactive proof protocol $(V,P)$ that fulfill the theorem based on Ref. \cite{Caro2024Classical}. The protocol is described as following:
    \begin{itemize}
        \item\textbf{Step 1: } Based on the algorithm in Theorem \ref{T noise1}, the verifier $V$ asks the prover $P$ to provide $k=O\left(\frac{\log(n/\delta)}{\tau^4}\right)$ samples from noisy QFS circuit, and produce a set $L$ containing $H=\{s\in\mc X_n:|\hat g(s)|\ge \tau\}=\{s\in\mc X_n:|\hat g(s)|\ne 0\}$ with probability at least $1-\frac{\delta}{3}$, using samples from $P$ and $k'_1=O\left(\frac{\log(1/\tau\delta)}{\tau^2}\right)$ samples from the random example oracle.
        \item\textbf{Step 2: }  The verifier $V$ checks the reliability of $L$ by obtaining a $\frac{1}{8}\tau^3$-approximation $\tilde{g}(s)$ of $\hat g(s)$ for $s\in L$ with probability $1-\frac{\delta}{3}$ using $k'_2=O\left(\frac{\log(1/\tau\delta)}{\tau^6}\right)$ samples from the random example oracle. If the sum $\tilde S=\sum_{s\in L}(\tilde g(s))^2\ge 1-\frac{1}{2}\tau^2$, the verifier chooses to trust $H\subseteq L$, otherwise it rejects the interaction.
        \item\textbf{Step 3: }  The verifier $V$ obtains an $\eps$-approximation $\tilde{g}'(s)$ of $\hat g(s)$ for $s\in L$ with probability $1-\frac{\delta}{3}$ using $k'_3=O\left(\frac{\log(1/\tau\delta)}{\eps^2}\right)$ samples from the random example oracle. Then $V$ chooses $s_0=\arg\max_{s\in L}\tilde g(s)$ and outputs $h(x)=s_0\cdot x$ as the learning result.
    \end{itemize}

    The method used in Step 2 and 3 to generate estimations of $\hat g(s)$ for $s\in L$ is similar to the one in Corollary \ref{T noise2}. In Step 2, $|\hat g(s)-\tilde{g}(s)|\le\frac{1}{8}\tau^3$ ensures that $|S-\tilde S|\le\frac{1}{2}\tau^2$ for $S=\sum_{s\in L}(\hat g(s))^2$. If $H\subseteq L$, then $S=1$ and  $\tilde S\ge1-\frac{1}{2}\tau^2$, otherwise $S\le1-\tau^2$ and $\tilde S<1-\frac{1}{2}\tau^2$. So with probability at least $1-\frac{\delta}{3}$, $L$ passes the validation if and only if $H\subseteq L$. For Step 3, if $H\subseteq L$, with probability at least $1-\frac{\delta}{3}$ we will find an $\eps$-approximation $\tilde{g}'(s)$ of $\hat g(s)$ for $s\in L$. Following the proof of Theorem \ref{T pac} it is straightforward to obtain that $V$ finds a correct answer for the learning task. For the proof of completeness: if $P$ is a honest quantum prover, then each step of the protocol successes with probability at least $1-\frac{\delta}{3}$, so the probability of $V$ returning a correct answer is at least $1-\delta$. For the proof of soundness: assume $L$ is already decided in Step 1, we consider the probability space over the samples in Step 2 and 3. Assuming $V$ gives a wrong answer, if $H\subseteq L$, then it means Step 3 fails, otherwise $H\not\subseteq L$ but $L$ passes the validation in Step 2. Both situation have probability no larger than $\frac{\delta}{3}$, so the probability of $V$ returning a incorrect answer is at most $\delta/3$.
\end{proof}

\textit{Discussion.}--- 
Our introduced quantum error-rectification algorithm can be extended to other QFS-based learning tasks, including agnostic improper Fourier-sparse learning, quantum disjunctive-normal-form  learning \cite{Bshouty1998Learning}, and their distributional extensions \cite{Caro2024Classical}. In this work, we have focused our discussion on the case of supervised learning. Yet, other learning scenarios, such as unsupervised learning and reinforcement learning \cite{Bengio2017Deep}, exist and their delegation to quantum servers would be highly desirable as well. Thus, it is of both fundamental and practical interest to extend the classical verification protocol to these scenarios. In addition, recent works have shown exponential separations between quantum algorithms with and without access to a quantum memory for certain problems \cite{huang2021information,Chen2022Exponential}. This gives rise to a natural question concerning whether these exponential separations carry over to the delegated noisy learning setting considered in this paper. 
Finally, it would be interesting and important to carry out an experiment to demonstrate the classical verification of quantum learning advantages on current noisy intermediate-scale quantum devices. Indeed, given that the verifier in our protocol is fully classical, with no requirements on capability of preparing single-qubit states and then teleporting them to the server as in blind quantum computing \cite{Barz2012Demonstration}, such an experiment is feasible on a verity of current quantum computing platforms, such as trapped ions \cite{Bruzewicz2019Trapped,Monore2021Programmable,Georgescu2020Trapped}, superconducting qubits \cite{Kjaergaard2020Superconducting,Arute2019Quantum, Ren2022Experimental}, and Ryderger atoms \cite{Saffman2010Quantum,Bluvstein2024Logical}. 
Demonstration of such an experiment would be a crucial step toward reliable delegation of learning tasks to untrusted noisy quantum servers that are currently available on cloud.

We acknowledge helpful discussions with Xilin Zheng, Haimeng Zhao, Weikang Li and Zhide Lu. 
This work was supported by the National Natural Science Foundation of China (Grants No. T2225008, and No. 12075128),  Shanghai Qi Zhi Institute,
the Innovation Program for Quantum Science and Technology (No. 2021ZD0302203), and Tsinghua University Dushi Program.

\bibliography{ref}

\begin{thebibliography}{60}%
\makeatletter
\providecommand \@ifxundefined [1]{%
 \@ifx{#1\undefined}
}%
\providecommand \@ifnum [1]{%
 \ifnum #1\expandafter \@firstoftwo
 \else \expandafter \@secondoftwo
 \fi
}%
\providecommand \@ifx [1]{%
 \ifx #1\expandafter \@firstoftwo
 \else \expandafter \@secondoftwo
 \fi
}%
\providecommand \natexlab [1]{#1}%
\providecommand \enquote  [1]{``#1''}%
\providecommand \bibnamefont  [1]{#1}%
\providecommand \bibfnamefont [1]{#1}%
\providecommand \citenamefont [1]{#1}%
\providecommand \href@noop [0]{\@secondoftwo}%
\providecommand \href [0]{\begingroup \@sanitize@url \@href}%
\providecommand \@href[1]{\@@startlink{#1}\@@href}%
\providecommand \@@href[1]{\endgroup#1\@@endlink}%
\providecommand \@sanitize@url [0]{\catcode `\\12\catcode `\$12\catcode `\&12\catcode `\#12\catcode `\^12\catcode `\_12\catcode `\%12\relax}%
\providecommand \@@startlink[1]{}%
\providecommand \@@endlink[0]{}%
\providecommand \url  [0]{\begingroup\@sanitize@url \@url }%
\providecommand \@url [1]{\endgroup\@href {#1}{\urlprefix }}%
\providecommand \urlprefix  [0]{URL }%
\providecommand \Eprint [0]{\href }%
\providecommand \doibase [0]{https://doi.org/}%
\providecommand \selectlanguage [0]{\@gobble}%
\providecommand \bibinfo  [0]{\@secondoftwo}%
\providecommand \bibfield  [0]{\@secondoftwo}%
\providecommand \translation [1]{[#1]}%
\providecommand \BibitemOpen [0]{}%
\providecommand \bibitemStop [0]{}%
\providecommand \bibitemNoStop [0]{.\EOS\space}%
\providecommand \EOS [0]{\spacefactor3000\relax}%
\providecommand \BibitemShut  [1]{\csname bibitem#1\endcsname}%
\let\auto@bib@innerbib\@empty
%</preamble>
\bibitem [{\citenamefont {Biamonte}\ \emph {et~al.}(2017)\citenamefont {Biamonte}, \citenamefont {Wittek}, \citenamefont {Pancotti}, \citenamefont {Rebentrost}, \citenamefont {Wiebe},\ and\ \citenamefont {Lloyd}}]{Biamonte2017Quantum}%
  \BibitemOpen
  \bibfield  {author} {\bibinfo {author} {\bibfnamefont {J.}~\bibnamefont {Biamonte}}, \bibinfo {author} {\bibfnamefont {P.}~\bibnamefont {Wittek}}, \bibinfo {author} {\bibfnamefont {N.}~\bibnamefont {Pancotti}}, \bibinfo {author} {\bibfnamefont {P.}~\bibnamefont {Rebentrost}}, \bibinfo {author} {\bibfnamefont {N.}~\bibnamefont {Wiebe}},\ and\ \bibinfo {author} {\bibfnamefont {S.}~\bibnamefont {Lloyd}},\ }\bibfield  {title} {\bibinfo {title} {Quantum machine learning},\ }\href {https://doi.org/10.1038/nature23474} {\bibfield  {journal} {\bibinfo  {journal} {Nature}\ }\textbf {\bibinfo {volume} {549}},\ \bibinfo {pages} {195} (\bibinfo {year} {2017})}\BibitemShut {NoStop}%
\bibitem [{\citenamefont {Das~Sarma}\ \emph {et~al.}(2019)\citenamefont {Das~Sarma}, \citenamefont {Deng},\ and\ \citenamefont {Duan}}]{DasSarma2019Machine}%
  \BibitemOpen
  \bibfield  {author} {\bibinfo {author} {\bibfnamefont {S.}~\bibnamefont {Das~Sarma}}, \bibinfo {author} {\bibfnamefont {D.-L.}\ \bibnamefont {Deng}},\ and\ \bibinfo {author} {\bibfnamefont {L.-M.}\ \bibnamefont {Duan}},\ }\bibfield  {title} {\bibinfo {title} {{Machine learning meets quantum physics}},\ }\href {https://doi.org/10.1063/PT.3.4164} {\bibfield  {journal} {\bibinfo  {journal} {Phys. Today}\ }\textbf {\bibinfo {volume} {72}},\ \bibinfo {pages} {48} (\bibinfo {year} {2019})}\BibitemShut {NoStop}%
\bibitem [{\citenamefont {Dunjko}\ and\ \citenamefont {Briegel}(2018)}]{Dunjko2018Machine}%
  \BibitemOpen
  \bibfield  {author} {\bibinfo {author} {\bibfnamefont {V.}~\bibnamefont {Dunjko}}\ and\ \bibinfo {author} {\bibfnamefont {H.~J.}\ \bibnamefont {Briegel}},\ }\bibfield  {title} {\bibinfo {title} {Machine learning \& artificial intelligence in the quantum domain: a review of recent progress},\ }\href {https://doi.org/10.1088/1361-6633/aab406} {\bibfield  {journal} {\bibinfo  {journal} {Rep. Prog. Phys.}\ }\textbf {\bibinfo {volume} {81}},\ \bibinfo {pages} {074001} (\bibinfo {year} {2018})}\BibitemShut {NoStop}%
\bibitem [{\citenamefont {Cerezo}\ \emph {et~al.}(2022)\citenamefont {Cerezo}, \citenamefont {Verdon}, \citenamefont {Huang}, \citenamefont {Cincio},\ and\ \citenamefont {Coles}}]{Cerezo2022Challenges}%
  \BibitemOpen
  \bibfield  {author} {\bibinfo {author} {\bibfnamefont {M.}~\bibnamefont {Cerezo}}, \bibinfo {author} {\bibfnamefont {G.}~\bibnamefont {Verdon}}, \bibinfo {author} {\bibfnamefont {H.-Y.}\ \bibnamefont {Huang}}, \bibinfo {author} {\bibfnamefont {L.}~\bibnamefont {Cincio}},\ and\ \bibinfo {author} {\bibfnamefont {P.~J.}\ \bibnamefont {Coles}},\ }\bibfield  {title} {\bibinfo {title} {Challenges and opportunities in quantum machine learning},\ }\href {https://doi.org/10.1038/s43588-022-00311-3} {\bibfield  {journal} {\bibinfo  {journal} {Nat. Comput. Sci.}\ }\textbf {\bibinfo {volume} {2}},\ \bibinfo {pages} {567} (\bibinfo {year} {2022})}\BibitemShut {NoStop}%
\bibitem [{\citenamefont {Lloyd}\ \emph {et~al.}(2014)\citenamefont {Lloyd}, \citenamefont {Mohseni},\ and\ \citenamefont {Rebentrost}}]{Lloyd2014Quantum}%
  \BibitemOpen
  \bibfield  {author} {\bibinfo {author} {\bibfnamefont {S.}~\bibnamefont {Lloyd}}, \bibinfo {author} {\bibfnamefont {M.}~\bibnamefont {Mohseni}},\ and\ \bibinfo {author} {\bibfnamefont {P.}~\bibnamefont {Rebentrost}},\ }\bibfield  {title} {\bibinfo {title} {Quantum principal component analysis},\ }\href {https://doi.org/10.1038/nphys3029} {\bibfield  {journal} {\bibinfo  {journal} {Nat. Phys.}\ }\textbf {\bibinfo {volume} {10}},\ \bibinfo {pages} {631} (\bibinfo {year} {2014})}\BibitemShut {NoStop}%
\bibitem [{\citenamefont {Rebentrost}\ \emph {et~al.}(2014)\citenamefont {Rebentrost}, \citenamefont {Mohseni},\ and\ \citenamefont {Lloyd}}]{Rebentrost2014Quantum}%
  \BibitemOpen
  \bibfield  {author} {\bibinfo {author} {\bibfnamefont {P.}~\bibnamefont {Rebentrost}}, \bibinfo {author} {\bibfnamefont {M.}~\bibnamefont {Mohseni}},\ and\ \bibinfo {author} {\bibfnamefont {S.}~\bibnamefont {Lloyd}},\ }\bibfield  {title} {\bibinfo {title} {Quantum support vector machine for big data classification},\ }\href {https://doi.org/10.1103/PhysRevLett.113.130503} {\bibfield  {journal} {\bibinfo  {journal} {Phys. Rev. Lett.}\ }\textbf {\bibinfo {volume} {113}},\ \bibinfo {pages} {130503} (\bibinfo {year} {2014})}\BibitemShut {NoStop}%
\bibitem [{\citenamefont {Gao}\ \emph {et~al.}(2018)\citenamefont {Gao}, \citenamefont {Zhang},\ and\ \citenamefont {Duan}}]{Gao2018Quantum}%
  \BibitemOpen
  \bibfield  {author} {\bibinfo {author} {\bibfnamefont {X.}~\bibnamefont {Gao}}, \bibinfo {author} {\bibfnamefont {Z.-Y.}\ \bibnamefont {Zhang}},\ and\ \bibinfo {author} {\bibfnamefont {L.-M.}\ \bibnamefont {Duan}},\ }\bibfield  {title} {\bibinfo {title} {A quantum machine learning algorithm based on generative models},\ }\href {https://doi.org/10.1126/sciadv.aat9004} {\bibfield  {journal} {\bibinfo  {journal} {Sci. Adv.}\ }\textbf {\bibinfo {volume} {4}},\ \bibinfo {pages} {eaat9004} (\bibinfo {year} {2018})}\BibitemShut {NoStop}%
\bibitem [{\citenamefont {Liu}\ \emph {et~al.}(2021)\citenamefont {Liu}, \citenamefont {Arunachalam},\ and\ \citenamefont {Temme}}]{Liu2021Rigorous}%
  \BibitemOpen
  \bibfield  {author} {\bibinfo {author} {\bibfnamefont {Y.}~\bibnamefont {Liu}}, \bibinfo {author} {\bibfnamefont {S.}~\bibnamefont {Arunachalam}},\ and\ \bibinfo {author} {\bibfnamefont {K.}~\bibnamefont {Temme}},\ }\bibfield  {title} {\bibinfo {title} {A rigorous and robust quantum speed-up in supervised machine learning},\ }\href {https://doi.org/10.1038/s41567-021-01287-z} {\bibfield  {journal} {\bibinfo  {journal} {Nat. Phys.}\ }\textbf {\bibinfo {volume} {17}},\ \bibinfo {pages} {1013} (\bibinfo {year} {2021})}\BibitemShut {NoStop}%
\bibitem [{\citenamefont {Grover}(1996)}]{Grover1996fast}%
  \BibitemOpen
  \bibfield  {author} {\bibinfo {author} {\bibfnamefont {L.~K.}\ \bibnamefont {Grover}},\ }\bibfield  {title} {\bibinfo {title} {A fast quantum mechanical algorithm for database search},\ }in\ \href {https://doi.org/10.1145/237814.237866} {\emph {\bibinfo {booktitle} {Proceedings of the Twenty-Eighth Annual {{ACM}} Symposium on {{Theory}} of Computing}}}\ (\bibinfo {year} {1996})\ pp.\ \bibinfo {pages} {212--219}\BibitemShut {NoStop}%
\bibitem [{\citenamefont {Brassard}\ \emph {et~al.}(2002)\citenamefont {Brassard}, \citenamefont {Hoyer}, \citenamefont {Mosca},\ and\ \citenamefont {Tapp}}]{Brassard2002Quantum}%
  \BibitemOpen
  \bibfield  {author} {\bibinfo {author} {\bibfnamefont {G.}~\bibnamefont {Brassard}}, \bibinfo {author} {\bibfnamefont {P.}~\bibnamefont {Hoyer}}, \bibinfo {author} {\bibfnamefont {M.}~\bibnamefont {Mosca}},\ and\ \bibinfo {author} {\bibfnamefont {A.}~\bibnamefont {Tapp}},\ }\bibfield  {title} {\bibinfo {title} {Quantum amplitude amplification and estimation},\ }\href {https://doi.org/10.1090/conm/305/05215} {\bibfield  {journal} {\bibinfo  {journal} {Contemp. Math.}\ }\textbf {\bibinfo {volume} {305}},\ \bibinfo {pages} {53} (\bibinfo {year} {2002})}\BibitemShut {NoStop}%
\bibitem [{\citenamefont {Harrow}\ \emph {et~al.}(2009)\citenamefont {Harrow}, \citenamefont {Hassidim},\ and\ \citenamefont {Lloyd}}]{Harrow2009Quantum}%
  \BibitemOpen
  \bibfield  {author} {\bibinfo {author} {\bibfnamefont {A.~W.}\ \bibnamefont {Harrow}}, \bibinfo {author} {\bibfnamefont {A.}~\bibnamefont {Hassidim}},\ and\ \bibinfo {author} {\bibfnamefont {S.}~\bibnamefont {Lloyd}},\ }\bibfield  {title} {\bibinfo {title} {Quantum algorithm for linear systems of equations},\ }\href {https://doi.org/10.1103/PhysRevLett.103.150502} {\bibfield  {journal} {\bibinfo  {journal} {Phys. Rev. Lett.}\ }\textbf {\bibinfo {volume} {103}},\ \bibinfo {pages} {150502} (\bibinfo {year} {2009})}\BibitemShut {NoStop}%
\bibitem [{\citenamefont {Valiant}(1984)}]{Valiant1984Theory}%
  \BibitemOpen
  \bibfield  {author} {\bibinfo {author} {\bibfnamefont {L.~G.}\ \bibnamefont {Valiant}},\ }\bibfield  {title} {\bibinfo {title} {A theory of the learnable},\ }\href {https://doi.org/10.1145/1968.1972} {\bibfield  {journal} {\bibinfo  {journal} {Commun. ACM}\ }\textbf {\bibinfo {volume} {27}},\ \bibinfo {pages} {1134} (\bibinfo {year} {1984})}\BibitemShut {NoStop}%
\bibitem [{\citenamefont {Haussler}(1992)}]{Haussler1992Decision}%
  \BibitemOpen
  \bibfield  {author} {\bibinfo {author} {\bibfnamefont {D.}~\bibnamefont {Haussler}},\ }\bibfield  {title} {\bibinfo {title} {Decision theoretic generalizations of the {{PAC}} model for neural net and other learning applications},\ }\href {https://doi.org/10.1016/0890-5401(92)90010-d} {\bibfield  {journal} {\bibinfo  {journal} {Inf. Comput.}\ }\textbf {\bibinfo {volume} {100}},\ \bibinfo {pages} {78} (\bibinfo {year} {1992})}\BibitemShut {NoStop}%
\bibitem [{\citenamefont {Kearns}\ \emph {et~al.}(1994)\citenamefont {Kearns}, \citenamefont {Schapire},\ and\ \citenamefont {Sellie}}]{Kearns1994Efficient}%
  \BibitemOpen
  \bibfield  {author} {\bibinfo {author} {\bibfnamefont {M.~J.}\ \bibnamefont {Kearns}}, \bibinfo {author} {\bibfnamefont {R.~E.}\ \bibnamefont {Schapire}},\ and\ \bibinfo {author} {\bibfnamefont {L.~M.}\ \bibnamefont {Sellie}},\ }\bibfield  {title} {\bibinfo {title} {Toward efficient agnostic learning},\ }\href {https://doi.org/10.1007/bf00993468} {\bibfield  {journal} {\bibinfo  {journal} {Mach. Learn.}\ }\textbf {\bibinfo {volume} {17}},\ \bibinfo {pages} {115} (\bibinfo {year} {1994})}\BibitemShut {NoStop}%
\bibitem [{\citenamefont {Bshouty}\ and\ \citenamefont {Jackson}(1998)}]{Bshouty1998Learning}%
  \BibitemOpen
  \bibfield  {author} {\bibinfo {author} {\bibfnamefont {N.~H.}\ \bibnamefont {Bshouty}}\ and\ \bibinfo {author} {\bibfnamefont {J.~C.}\ \bibnamefont {Jackson}},\ }\bibfield  {title} {\bibinfo {title} {Learning {{DNF}} over the {{Uniform Distribution Using}} a {{Quantum Example Oracle}}},\ }\href {https://doi.org/10.1137/s0097539795293123} {\bibfield  {journal} {\bibinfo  {journal} {SIAM J. Comput.}\ }\textbf {\bibinfo {volume} {28}},\ \bibinfo {pages} {1136} (\bibinfo {year} {1998})}\BibitemShut {NoStop}%
\bibitem [{\citenamefont {At{\i}c{\i}}\ and\ \citenamefont {Servedio}(2007)}]{Atici2007Quantum}%
  \BibitemOpen
  \bibfield  {author} {\bibinfo {author} {\bibfnamefont {A.}~\bibnamefont {At{\i}c{\i}}}\ and\ \bibinfo {author} {\bibfnamefont {R.~A.}\ \bibnamefont {Servedio}},\ }\bibfield  {title} {\bibinfo {title} {Quantum algorithms for learning and testing juntas},\ }\href {https://doi.org/10.1007/s11128-007-0061-6} {\bibfield  {journal} {\bibinfo  {journal} {Quantum Inf. Process.}\ }\textbf {\bibinfo {volume} {6}},\ \bibinfo {pages} {323} (\bibinfo {year} {2007})}\BibitemShut {NoStop}%
\bibitem [{\citenamefont {Servedio}\ and\ \citenamefont {Gortler}(2004)}]{Servedio2004Equivalences}%
  \BibitemOpen
  \bibfield  {author} {\bibinfo {author} {\bibfnamefont {R.~A.}\ \bibnamefont {Servedio}}\ and\ \bibinfo {author} {\bibfnamefont {S.~J.}\ \bibnamefont {Gortler}},\ }\bibfield  {title} {\bibinfo {title} {Equivalences and {{Separations Between Quantum}} and {{Classical Learnability}}},\ }\href {https://doi.org/10.1137/s0097539704412910} {\bibfield  {journal} {\bibinfo  {journal} {SIAM J. Comput.}\ }\textbf {\bibinfo {volume} {33}},\ \bibinfo {pages} {1067} (\bibinfo {year} {2004})}\BibitemShut {NoStop}%
\bibitem [{\citenamefont {Montanaro}(2012)}]{Montanaro2012quantum}%
  \BibitemOpen
  \bibfield  {author} {\bibinfo {author} {\bibfnamefont {A.}~\bibnamefont {Montanaro}},\ }\bibfield  {title} {\bibinfo {title} {The quantum query complexity of learning multilinear polynomials},\ }\href {https://doi.org/10.1016/j.ipl.2012.03.002} {\bibfield  {journal} {\bibinfo  {journal} {Inf. Process. Lett.}\ }\textbf {\bibinfo {volume} {112}},\ \bibinfo {pages} {438} (\bibinfo {year} {2012})}\BibitemShut {NoStop}%
\bibitem [{\citenamefont {Arunachalam}\ \emph {et~al.}(2021)\citenamefont {Arunachalam}, \citenamefont {Chakraborty}, \citenamefont {Lee}, \citenamefont {Paraashar},\ and\ \citenamefont {{de Wolf}}}]{Arunachalam2021Two}%
  \BibitemOpen
  \bibfield  {author} {\bibinfo {author} {\bibfnamefont {S.}~\bibnamefont {Arunachalam}}, \bibinfo {author} {\bibfnamefont {S.}~\bibnamefont {Chakraborty}}, \bibinfo {author} {\bibfnamefont {T.}~\bibnamefont {Lee}}, \bibinfo {author} {\bibfnamefont {M.}~\bibnamefont {Paraashar}},\ and\ \bibinfo {author} {\bibfnamefont {R.}~\bibnamefont {{de Wolf}}},\ }\bibfield  {title} {\bibinfo {title} {Two new results about quantum exact learning},\ }\href {https://doi.org/10.22331/q-2021-11-24-587} {\bibfield  {journal} {\bibinfo  {journal} {Quantum}\ }\textbf {\bibinfo {volume} {5}},\ \bibinfo {pages} {587} (\bibinfo {year} {2021})}\BibitemShut {NoStop}%
\bibitem [{\citenamefont {Caro}\ \emph {et~al.}(2024)\citenamefont {Caro}, \citenamefont {Hinsche}, \citenamefont {Ioannou}, \citenamefont {Nietner},\ and\ \citenamefont {Sweke}}]{Caro2024Classical}%
  \BibitemOpen
  \bibfield  {author} {\bibinfo {author} {\bibfnamefont {M.~C.}\ \bibnamefont {Caro}}, \bibinfo {author} {\bibfnamefont {M.}~\bibnamefont {Hinsche}}, \bibinfo {author} {\bibfnamefont {M.}~\bibnamefont {Ioannou}}, \bibinfo {author} {\bibfnamefont {A.}~\bibnamefont {Nietner}},\ and\ \bibinfo {author} {\bibfnamefont {R.}~\bibnamefont {Sweke}},\ }\bibfield  {title} {\bibinfo {title} {{Classical Verification of Quantum Learning}},\ }in\ \href {https://doi.org/10.4230/LIPIcs.ITCS.2024.24} {\emph {\bibinfo {booktitle} {15th Innovations in Theoretical Computer Science Conference (ITCS 2024)}}},\ \bibinfo {series} {Leibniz International Proceedings in Informatics (LIPIcs)}, Vol.\ \bibinfo {volume} {287}\ (\bibinfo  {publisher} {Schloss Dagstuhl -- Leibniz-Zentrum f{\"u}r Informatik},\ \bibinfo {address} {Dagstuhl, Germany},\ \bibinfo {year} {2024})\ pp.\ \bibinfo {pages} {24:1--24:23}\BibitemShut {NoStop}%
\bibitem [{\citenamefont {Arunachalam}\ and\ \citenamefont {De~Wolf}(2018)}]{Arunachalam2018Optimal}%
  \BibitemOpen
  \bibfield  {author} {\bibinfo {author} {\bibfnamefont {S.}~\bibnamefont {Arunachalam}}\ and\ \bibinfo {author} {\bibfnamefont {R.}~\bibnamefont {De~Wolf}},\ }\bibfield  {title} {\bibinfo {title} {Optimal quantum sample complexity of learning algorithms},\ }\href {http://jmlr.org/papers/v19/18-195.html} {\bibfield  {journal} {\bibinfo  {journal} {J. Mach. Learn. Res.}\ }\textbf {\bibinfo {volume} {19}},\ \bibinfo {pages} {1} (\bibinfo {year} {2018})}\BibitemShut {NoStop}%
\bibitem [{\citenamefont {Atici}\ and\ \citenamefont {Servedio}(2005)}]{Atici2005Improved}%
  \BibitemOpen
  \bibfield  {author} {\bibinfo {author} {\bibfnamefont {A.}~\bibnamefont {Atici}}\ and\ \bibinfo {author} {\bibfnamefont {R.~A.}\ \bibnamefont {Servedio}},\ }\bibfield  {title} {\bibinfo {title} {Improved {{Bounds}} on {{Quantum Learning Algorithms}}},\ }\href {https://doi.org/10.1007/s11128-005-0001-2} {\bibfield  {journal} {\bibinfo  {journal} {Quantum Inf. Process.}\ }\textbf {\bibinfo {volume} {4}},\ \bibinfo {pages} {355} (\bibinfo {year} {2005})}\BibitemShut {NoStop}%
\bibitem [{\citenamefont {Zhang}(2010)}]{Zhang2010improved}%
  \BibitemOpen
  \bibfield  {author} {\bibinfo {author} {\bibfnamefont {C.}~\bibnamefont {Zhang}},\ }\bibfield  {title} {\bibinfo {title} {An improved lower bound on query complexity for quantum {{PAC}} learning},\ }\href {https://doi.org/10.1016/j.ipl.2010.10.007} {\bibfield  {journal} {\bibinfo  {journal} {Inf. Process. Lett.}\ }\textbf {\bibinfo {volume} {111}},\ \bibinfo {pages} {40} (\bibinfo {year} {2010})}\BibitemShut {NoStop}%
\bibitem [{\citenamefont {Cross}\ \emph {et~al.}(2015)\citenamefont {Cross}, \citenamefont {Smith},\ and\ \citenamefont {Smolin}}]{Cross2015Quantum}%
  \BibitemOpen
  \bibfield  {author} {\bibinfo {author} {\bibfnamefont {A.~W.}\ \bibnamefont {Cross}}, \bibinfo {author} {\bibfnamefont {G.}~\bibnamefont {Smith}},\ and\ \bibinfo {author} {\bibfnamefont {J.~A.}\ \bibnamefont {Smolin}},\ }\bibfield  {title} {\bibinfo {title} {Quantum learning robust against noise},\ }\href {https://doi.org/10.1103/PhysRevA.92.012327} {\bibfield  {journal} {\bibinfo  {journal} {Phys. Rev. A}\ }\textbf {\bibinfo {volume} {92}},\ \bibinfo {pages} {012327} (\bibinfo {year} {2015})}\BibitemShut {NoStop}%
\bibitem [{\citenamefont {Grilo}\ \emph {et~al.}(2019)\citenamefont {Grilo}, \citenamefont {Kerenidis},\ and\ \citenamefont {Zijlstra}}]{Grilo2019Learning}%
  \BibitemOpen
  \bibfield  {author} {\bibinfo {author} {\bibfnamefont {A.~B.}\ \bibnamefont {Grilo}}, \bibinfo {author} {\bibfnamefont {I.}~\bibnamefont {Kerenidis}},\ and\ \bibinfo {author} {\bibfnamefont {T.}~\bibnamefont {Zijlstra}},\ }\bibfield  {title} {\bibinfo {title} {Learning-with-errors problem is easy with quantum samples},\ }\href {https://doi.org/10.1103/PhysRevA.99.032314} {\bibfield  {journal} {\bibinfo  {journal} {Phys. Rev. A}\ }\textbf {\bibinfo {volume} {99}},\ \bibinfo {pages} {032314} (\bibinfo {year} {2019})}\BibitemShut {NoStop}%
\bibitem [{\citenamefont {Bernstein}\ and\ \citenamefont {Vazirani}(1993)}]{Bernstein1993Quantum}%
  \BibitemOpen
  \bibfield  {author} {\bibinfo {author} {\bibfnamefont {E.}~\bibnamefont {Bernstein}}\ and\ \bibinfo {author} {\bibfnamefont {U.}~\bibnamefont {Vazirani}},\ }\bibfield  {title} {\bibinfo {title} {Quantum complexity theory},\ }in\ \href {https://doi.org/10.1145/167088.167097} {\emph {\bibinfo {booktitle} {Proceedings of the Twenty-Fifth Annual {{ACM}} Symposium on {{Theory}} of Computing - {{STOC}} '93}}},\ \bibinfo {series and number} {{{STOC}} '93}\ (\bibinfo  {publisher} {ACM Press},\ \bibinfo {year} {1993})\ pp.\ \bibinfo {pages} {11--20}\BibitemShut {NoStop}%
\bibitem [{\citenamefont {Shor}(1995)}]{Shor1995Scheme}%
  \BibitemOpen
  \bibfield  {author} {\bibinfo {author} {\bibfnamefont {P.~W.}\ \bibnamefont {Shor}},\ }\bibfield  {title} {\bibinfo {title} {Scheme for reducing decoherence in quantum computer memory},\ }\href {https://doi.org/10.1103/PhysRevA.52.R2493} {\bibfield  {journal} {\bibinfo  {journal} {Phys. Rev. A}\ }\textbf {\bibinfo {volume} {52}},\ \bibinfo {pages} {R2493} (\bibinfo {year} {1995})}\BibitemShut {NoStop}%
\bibitem [{\citenamefont {Steane}(1996)}]{Steane1996Simple}%
  \BibitemOpen
  \bibfield  {author} {\bibinfo {author} {\bibfnamefont {A.~M.}\ \bibnamefont {Steane}},\ }\bibfield  {title} {\bibinfo {title} {Simple quantum error-correcting codes},\ }\href {https://doi.org/10.1103/PhysRevA.54.4741} {\bibfield  {journal} {\bibinfo  {journal} {Phys. Rev. A}\ }\textbf {\bibinfo {volume} {54}},\ \bibinfo {pages} {4741} (\bibinfo {year} {1996})}\BibitemShut {NoStop}%
\bibitem [{\citenamefont {Knill}\ \emph {et~al.}(1998)\citenamefont {Knill}, \citenamefont {Laflamme},\ and\ \citenamefont {Zurek}}]{Knill1998Resilient}%
  \BibitemOpen
  \bibfield  {author} {\bibinfo {author} {\bibfnamefont {E.}~\bibnamefont {Knill}}, \bibinfo {author} {\bibfnamefont {R.}~\bibnamefont {Laflamme}},\ and\ \bibinfo {author} {\bibfnamefont {W.~H.}\ \bibnamefont {Zurek}},\ }\bibfield  {title} {\bibinfo {title} {Resilient quantum computation},\ }\href {https://doi.org/10.1126/science.279.5349.342} {\bibfield  {journal} {\bibinfo  {journal} {Science}\ }\textbf {\bibinfo {volume} {279}},\ \bibinfo {pages} {342} (\bibinfo {year} {1998})}\BibitemShut {NoStop}%
\bibitem [{\citenamefont {Terhal}(2015)}]{Terhal2015Quantum}%
  \BibitemOpen
  \bibfield  {author} {\bibinfo {author} {\bibfnamefont {B.~M.}\ \bibnamefont {Terhal}},\ }\bibfield  {title} {\bibinfo {title} {Quantum error correction for quantum memories},\ }\href {https://doi.org/10.1103/revmodphys.87.307} {\bibfield  {journal} {\bibinfo  {journal} {Rev. Mod. Phys.}\ }\textbf {\bibinfo {volume} {87}},\ \bibinfo {pages} {307} (\bibinfo {year} {2015})}\BibitemShut {NoStop}%
\bibitem [{\citenamefont {Preskill}(2018)}]{Preskill2018Quantum}%
  \BibitemOpen
  \bibfield  {author} {\bibinfo {author} {\bibfnamefont {J.}~\bibnamefont {Preskill}},\ }\bibfield  {title} {\bibinfo {title} {Quantum {{Computing}} in the {{NISQ}} era and beyond},\ }\href {https://doi.org/10.22331/q-2018-08-06-79} {\bibfield  {journal} {\bibinfo  {journal} {Quantum}\ }\textbf {\bibinfo {volume} {2}},\ \bibinfo {pages} {79} (\bibinfo {year} {2018})}\BibitemShut {NoStop}%
\bibitem [{202(2024{\natexlab{a}})}]{2024IBM}%
  \BibitemOpen
  \href {https://www.ibm.com/quantum/technology} {\bibinfo {title} {https://www.ibm.com/quantum/technology}} (\bibinfo {year} {2024}{\natexlab{a}})\BibitemShut {NoStop}%
\bibitem [{\citenamefont {Wurtz}\ \emph {et~al.}(2023)\citenamefont {Wurtz}, \citenamefont {Bylinskii}, \citenamefont {Braverman}, \citenamefont {Amato-Grill}, \citenamefont {Cantu}, \citenamefont {Huber}, \citenamefont {Lukin}, \citenamefont {Liu}, \citenamefont {Weinberg}, \citenamefont {Long}, \citenamefont {Wang}, \citenamefont {Gemelke},\ and\ \citenamefont {Keesling}}]{Wurtz2023Aquila}%
  \BibitemOpen
  \bibfield  {author} {\bibinfo {author} {\bibfnamefont {J.}~\bibnamefont {Wurtz}}, \bibinfo {author} {\bibfnamefont {A.}~\bibnamefont {Bylinskii}}, \bibinfo {author} {\bibfnamefont {B.}~\bibnamefont {Braverman}}, \bibinfo {author} {\bibfnamefont {J.}~\bibnamefont {Amato-Grill}}, \bibinfo {author} {\bibfnamefont {S.~H.}\ \bibnamefont {Cantu}}, \bibinfo {author} {\bibfnamefont {F.}~\bibnamefont {Huber}}, \bibinfo {author} {\bibfnamefont {A.}~\bibnamefont {Lukin}}, \bibinfo {author} {\bibfnamefont {F.}~\bibnamefont {Liu}}, \bibinfo {author} {\bibfnamefont {P.}~\bibnamefont {Weinberg}}, \bibinfo {author} {\bibfnamefont {J.}~\bibnamefont {Long}}, \bibinfo {author} {\bibfnamefont {S.-T.}\ \bibnamefont {Wang}}, \bibinfo {author} {\bibfnamefont {N.}~\bibnamefont {Gemelke}},\ and\ \bibinfo {author} {\bibfnamefont {A.}~\bibnamefont {Keesling}},\ }\href {https://arxiv.org/abs/2306.11727} {\bibinfo {title} {Aquila: Quera's 256-qubit neutral-atom quantum computer}} (\bibinfo {year} {2023}),\ \Eprint
  {https://arxiv.org/abs/2306.11727} {arXiv:2306.11727 [quant-ph]} \BibitemShut {NoStop}%
\bibitem [{202(2024{\natexlab{b}})}]{2024QMware}%
  \BibitemOpen
  \href {{https://www.qm-ware.com/product/qmware-cloud-platform/}} {\bibinfo {title} {https://www.qm-ware.com/product/qmware-cloud-platform/}} (\bibinfo {year} {2024}{\natexlab{b}})\BibitemShut {NoStop}%
\bibitem [{202(2024{\natexlab{c}})}]{2024Quafu}%
  \BibitemOpen
  \href {https://quafu.baqis.ac.cn/} {\bibinfo {title} {https://quafu.baqis.ac.cn/}} (\bibinfo {year} {2024}{\natexlab{c}})\BibitemShut {NoStop}%
\bibitem [{\citenamefont {Gheorghiu}\ \emph {et~al.}(2018)\citenamefont {Gheorghiu}, \citenamefont {Kapourniotis},\ and\ \citenamefont {Kashefi}}]{Gheorghiu2018Verification}%
  \BibitemOpen
  \bibfield  {author} {\bibinfo {author} {\bibfnamefont {A.}~\bibnamefont {Gheorghiu}}, \bibinfo {author} {\bibfnamefont {T.}~\bibnamefont {Kapourniotis}},\ and\ \bibinfo {author} {\bibfnamefont {E.}~\bibnamefont {Kashefi}},\ }\bibfield  {title} {\bibinfo {title} {Verification of {{Quantum Computation}}: {{An Overview}} of {{Existing Approaches}}},\ }\href {https://doi.org/10.1007/s00224-018-9872-3} {\bibfield  {journal} {\bibinfo  {journal} {Theory Comput. Syst.}\ }\textbf {\bibinfo {volume} {63}},\ \bibinfo {pages} {715} (\bibinfo {year} {2018})}\BibitemShut {NoStop}%
\bibitem [{\citenamefont {Mahadev}(2018)}]{Mahadev2018Classical}%
  \BibitemOpen
  \bibfield  {author} {\bibinfo {author} {\bibfnamefont {U.}~\bibnamefont {Mahadev}},\ }\bibfield  {title} {\bibinfo {title} {Classical {{Verification}} of {{Quantum Computations}}},\ }in\ \href {https://doi.org/10.1109/focs.2018.00033} {\emph {\bibinfo {booktitle} {2018 {{IEEE}} 59th {{Annual Symposium}} on {{Foundations}} of {{Computer Science}} ({{FOCS}})}}}\ (\bibinfo  {publisher} {IEEE},\ \bibinfo {year} {2018})\ pp.\ \bibinfo {pages} {259--267}\BibitemShut {NoStop}%
\bibitem [{\citenamefont {Fitzsimons}(2017)}]{Fitzsimons2017Private}%
  \BibitemOpen
  \bibfield  {author} {\bibinfo {author} {\bibfnamefont {J.~F.}\ \bibnamefont {Fitzsimons}},\ }\bibfield  {title} {\bibinfo {title} {Private quantum computation: an introduction to blind quantum computing and related protocols},\ }\href {https://doi.org/10.1038/s41534-017-0025-3} {\bibfield  {journal} {\bibinfo  {journal} {npj Quantum Inf.}\ }\textbf {\bibinfo {volume} {3}},\ \bibinfo {pages} {23} (\bibinfo {year} {2017})}\BibitemShut {NoStop}%
\bibitem [{\citenamefont {Broadbent}\ \emph {et~al.}(2009)\citenamefont {Broadbent}, \citenamefont {Fitzsimons},\ and\ \citenamefont {Kashefi}}]{Broadbent2009Universal}%
  \BibitemOpen
  \bibfield  {author} {\bibinfo {author} {\bibfnamefont {A.}~\bibnamefont {Broadbent}}, \bibinfo {author} {\bibfnamefont {J.}~\bibnamefont {Fitzsimons}},\ and\ \bibinfo {author} {\bibfnamefont {E.}~\bibnamefont {Kashefi}},\ }\bibfield  {title} {\bibinfo {title} {Universal {{Blind Quantum Computation}}},\ }in\ \href {https://doi.org/10.1109/focs.2009.36} {\emph {\bibinfo {booktitle} {2009 50th {{Annual IEEE Symposium}} on {{Foundations}} of {{Computer Science}}}}}\ (\bibinfo  {publisher} {IEEE},\ \bibinfo {year} {2009})\ pp.\ \bibinfo {pages} {517--526}\BibitemShut {NoStop}%
\bibitem [{\citenamefont {Goldwasser}\ \emph {et~al.}(2019)\citenamefont {Goldwasser}, \citenamefont {Micali},\ and\ \citenamefont {Rackoff}}]{Goldwasser2019knowledge}%
  \BibitemOpen
  \bibfield  {author} {\bibinfo {author} {\bibfnamefont {S.}~\bibnamefont {Goldwasser}}, \bibinfo {author} {\bibfnamefont {S.}~\bibnamefont {Micali}},\ and\ \bibinfo {author} {\bibfnamefont {C.}~\bibnamefont {Rackoff}},\ }\bibfield  {title} {\bibinfo {title} {The knowledge complexity of interactive proof-systems},\ }\href {https://doi.org/10.1145/3335741.3335750} {\bibfield  {journal} {\bibinfo  {journal} {Providing Sound Foundations for Cryptography: On the Work of Shafi Goldwasser and Silvio Micali}\ ,\ \bibinfo {pages} {203}} (\bibinfo {year} {2019})}\BibitemShut {NoStop}%
\bibitem [{\citenamefont {Shamir}(1992)}]{Shamir1992IP}%
  \BibitemOpen
  \bibfield  {author} {\bibinfo {author} {\bibfnamefont {A.}~\bibnamefont {Shamir}},\ }\bibfield  {title} {\bibinfo {title} {{{IP}} = {{PSPACE}}},\ }\href {https://doi.org/10.1145/146585.146609} {\bibfield  {journal} {\bibinfo  {journal} {J. ACM}\ }\textbf {\bibinfo {volume} {39}},\ \bibinfo {pages} {869} (\bibinfo {year} {1992})}\BibitemShut {NoStop}%
\bibitem [{\citenamefont {Goldwasser}\ \emph {et~al.}(2021)\citenamefont {Goldwasser}, \citenamefont {Rothblum}, \citenamefont {Shafer},\ and\ \citenamefont {Yehudayoff}}]{Goldwasser2021Interactive}%
  \BibitemOpen
  \bibfield  {author} {\bibinfo {author} {\bibfnamefont {S.}~\bibnamefont {Goldwasser}}, \bibinfo {author} {\bibfnamefont {G.~N.}\ \bibnamefont {Rothblum}}, \bibinfo {author} {\bibfnamefont {J.}~\bibnamefont {Shafer}},\ and\ \bibinfo {author} {\bibfnamefont {A.}~\bibnamefont {Yehudayoff}},\ }\bibfield  {title} {\bibinfo {title} {Interactive {{Proofs}} for {{Verifying Machine Learning}}},\ }in\ \href {https://doi.org/10.4230/LIPICS.ITCS.2021.41} {\emph {\bibinfo {booktitle} {12th Innovations in Theoretical Computer Science Conference (ITCS 2021)}}}\ (\bibinfo  {publisher} {Schloss Dagstuhl -- Leibniz-Zentrum f{\"u}r Informatik},\ \bibinfo {year} {2021})\BibitemShut {NoStop}%
\bibitem [{\citenamefont {Nielsen}\ and\ \citenamefont {Chuang}(2010)}]{Nielsen2010Quantum}%
  \BibitemOpen
  \bibfield  {author} {\bibinfo {author} {\bibfnamefont {M.~A.}\ \bibnamefont {Nielsen}}\ and\ \bibinfo {author} {\bibfnamefont {I.~L.}\ \bibnamefont {Chuang}},\ }\href@noop {} {\emph {\bibinfo {title} {Quantum Computation and Quantum Information}}}\ (\bibinfo  {publisher} {Cambridge university press},\ \bibinfo {year} {2010})\BibitemShut {NoStop}%
\bibitem [{\citenamefont {Angluin}(1988)}]{Angluin1988Queries}%
  \BibitemOpen
  \bibfield  {author} {\bibinfo {author} {\bibfnamefont {D.}~\bibnamefont {Angluin}},\ }\bibfield  {title} {\bibinfo {title} {Queries and concept learning},\ }\href {https://doi.org/10.1023/A:1022821128753} {\bibfield  {journal} {\bibinfo  {journal} {Mach. Learn.}\ }\textbf {\bibinfo {volume} {2}},\ \bibinfo {pages} {319} (\bibinfo {year} {1988})}\BibitemShut {NoStop}%
\bibitem [{\citenamefont {Goldreich}\ and\ \citenamefont {Levin}(1989)}]{Goldreich1989Hardcore}%
  \BibitemOpen
  \bibfield  {author} {\bibinfo {author} {\bibfnamefont {O.}~\bibnamefont {Goldreich}}\ and\ \bibinfo {author} {\bibfnamefont {L.~A.}\ \bibnamefont {Levin}},\ }\bibfield  {title} {\bibinfo {title} {A hard-core predicate for all one-way functions},\ }in\ \href {https://doi.org/10.1145/73007.73010} {\emph {\bibinfo {booktitle} {Proceedings of the Twenty-First Annual {{ACM}} Symposium on {{Theory}} of Computing}}},\ \bibinfo {series and number} {{{STOC}} '89}\ (\bibinfo  {publisher} {Association for Computing Machinery},\ \bibinfo {address} {New York, NY, USA},\ \bibinfo {year} {1989})\ pp.\ \bibinfo {pages} {25--32}\BibitemShut {NoStop}%
\bibitem [{\citenamefont {Kushilevitz}\ and\ \citenamefont {Mansour}(1991)}]{Kushilevitz1991Learning}%
  \BibitemOpen
  \bibfield  {author} {\bibinfo {author} {\bibfnamefont {E.}~\bibnamefont {Kushilevitz}}\ and\ \bibinfo {author} {\bibfnamefont {Y.}~\bibnamefont {Mansour}},\ }\bibfield  {title} {\bibinfo {title} {Learning decision trees using the {{Fourier}} spectrum},\ }in\ \href {https://doi.org/10.1145/103418.103466} {\emph {\bibinfo {booktitle} {Proceedings of the Twenty-Third Annual {{ACM}} Symposium on {{Theory}} of {{Computing}}}}},\ \bibinfo {series and number} {{{STOC}} '91}\ (\bibinfo  {publisher} {Association for Computing Machinery},\ \bibinfo {address} {New York, NY, USA},\ \bibinfo {year} {1991})\ pp.\ \bibinfo {pages} {455--464}\BibitemShut {NoStop}%
\bibitem [{\citenamefont {Lyubashevsky}(2005)}]{Lyubashevsky2005Parity}%
  \BibitemOpen
  \bibfield  {author} {\bibinfo {author} {\bibfnamefont {V.}~\bibnamefont {Lyubashevsky}},\ }\bibfield  {title} {\bibinfo {title} {The parity problem in the presence of noise, decoding random linear codes, and the subset sum problem},\ }in\ \href {https://doi.org/10.1007/11538462_32} {\emph {\bibinfo {booktitle} {International Workshop on Approximation Algorithms for Combinatorial Optimization}}}\ (\bibinfo {organization} {Springer},\ \bibinfo {year} {2005})\ pp.\ \bibinfo {pages} {378--389}\BibitemShut {NoStop}%
\bibitem [{\citenamefont {Regev}(2009)}]{Regev2009Lattices}%
  \BibitemOpen
  \bibfield  {author} {\bibinfo {author} {\bibfnamefont {O.}~\bibnamefont {Regev}},\ }\bibfield  {title} {\bibinfo {title} {On lattices, learning with errors, random linear codes, and cryptography},\ }\href {https://doi.org/10.1145/1568318.1568324} {\bibfield  {journal} {\bibinfo  {journal} {J. ACM}\ }\textbf {\bibinfo {volume} {56}},\ \bibinfo {pages} {1} (\bibinfo {year} {2009})}\BibitemShut {NoStop}%
\bibitem [{\citenamefont {Bengio}\ \emph {et~al.}(2017)\citenamefont {Bengio}, \citenamefont {Goodfellow},\ and\ \citenamefont {Courville}}]{Bengio2017Deep}%
  \BibitemOpen
  \bibfield  {author} {\bibinfo {author} {\bibfnamefont {Y.}~\bibnamefont {Bengio}}, \bibinfo {author} {\bibfnamefont {I.}~\bibnamefont {Goodfellow}},\ and\ \bibinfo {author} {\bibfnamefont {A.}~\bibnamefont {Courville}},\ }\href@noop {} {\emph {\bibinfo {title} {Deep learning}}},\ Vol.~\bibinfo {volume} {1}\ (\bibinfo  {publisher} {MIT press Cambridge, MA, USA},\ \bibinfo {year} {2017})\BibitemShut {NoStop}%
\bibitem [{\citenamefont {Huang}\ \emph {et~al.}(2021)\citenamefont {Huang}, \citenamefont {Kueng},\ and\ \citenamefont {Preskill}}]{huang2021information}%
  \BibitemOpen
  \bibfield  {author} {\bibinfo {author} {\bibfnamefont {H.-Y.}\ \bibnamefont {Huang}}, \bibinfo {author} {\bibfnamefont {R.}~\bibnamefont {Kueng}},\ and\ \bibinfo {author} {\bibfnamefont {J.}~\bibnamefont {Preskill}},\ }\bibfield  {title} {\bibinfo {title} {Information-theoretic bounds on quantum advantage in machine learning},\ }\href {https://doi.org/10.1103/PhysRevLett.126.190505} {\bibfield  {journal} {\bibinfo  {journal} {Phys. Rev. Lett.}\ }\textbf {\bibinfo {volume} {126}},\ \bibinfo {pages} {190505} (\bibinfo {year} {2021})}\BibitemShut {NoStop}%
\bibitem [{\citenamefont {Chen}\ \emph {et~al.}(2022)\citenamefont {Chen}, \citenamefont {Cotler}, \citenamefont {Huang},\ and\ \citenamefont {Li}}]{Chen2022Exponential}%
  \BibitemOpen
  \bibfield  {author} {\bibinfo {author} {\bibfnamefont {S.}~\bibnamefont {Chen}}, \bibinfo {author} {\bibfnamefont {J.}~\bibnamefont {Cotler}}, \bibinfo {author} {\bibfnamefont {H.-Y.}\ \bibnamefont {Huang}},\ and\ \bibinfo {author} {\bibfnamefont {J.}~\bibnamefont {Li}},\ }\bibfield  {title} {\bibinfo {title} {Exponential separations between learning with and without quantum memory},\ }in\ \href {https://doi.org/10.1109/FOCS52979.2021.00063} {\emph {\bibinfo {booktitle} {2021 IEEE 62nd Annual Symposium on Foundations of Computer Science (FOCS)}}}\ (\bibinfo {organization} {IEEE},\ \bibinfo {year} {2022})\ pp.\ \bibinfo {pages} {574--585}\BibitemShut {NoStop}%
\bibitem [{\citenamefont {Barz}\ \emph {et~al.}(2012)\citenamefont {Barz}, \citenamefont {Kashefi}, \citenamefont {Broadbent}, \citenamefont {Fitzsimons}, \citenamefont {Zeilinger},\ and\ \citenamefont {Walther}}]{Barz2012Demonstration}%
  \BibitemOpen
  \bibfield  {author} {\bibinfo {author} {\bibfnamefont {S.}~\bibnamefont {Barz}}, \bibinfo {author} {\bibfnamefont {E.}~\bibnamefont {Kashefi}}, \bibinfo {author} {\bibfnamefont {A.}~\bibnamefont {Broadbent}}, \bibinfo {author} {\bibfnamefont {J.~F.}\ \bibnamefont {Fitzsimons}}, \bibinfo {author} {\bibfnamefont {A.}~\bibnamefont {Zeilinger}},\ and\ \bibinfo {author} {\bibfnamefont {P.}~\bibnamefont {Walther}},\ }\bibfield  {title} {\bibinfo {title} {Demonstration of blind quantum computing},\ }\href {https://www.science.org/doi/abs/10.1126/science.1214707} {\bibfield  {journal} {\bibinfo  {journal} {Science}\ }\textbf {\bibinfo {volume} {335}},\ \bibinfo {pages} {303} (\bibinfo {year} {2012})}\BibitemShut {NoStop}%
\bibitem [{\citenamefont {Bruzewicz}\ \emph {et~al.}(2019)\citenamefont {Bruzewicz}, \citenamefont {Chiaverini}, \citenamefont {McConnell},\ and\ \citenamefont {Sage}}]{Bruzewicz2019Trapped}%
  \BibitemOpen
  \bibfield  {author} {\bibinfo {author} {\bibfnamefont {C.~D.}\ \bibnamefont {Bruzewicz}}, \bibinfo {author} {\bibfnamefont {J.}~\bibnamefont {Chiaverini}}, \bibinfo {author} {\bibfnamefont {R.}~\bibnamefont {McConnell}},\ and\ \bibinfo {author} {\bibfnamefont {J.~M.}\ \bibnamefont {Sage}},\ }\bibfield  {title} {\bibinfo {title} {Trapped-ion quantum computing: Progress and challenges},\ }\href {https://doi.org/10.1063/1.5088164} {\bibfield  {journal} {\bibinfo  {journal} {Appl. Phys. Rev.}\ }\textbf {\bibinfo {volume} {6}} (\bibinfo {year} {2019})}\BibitemShut {NoStop}%
\bibitem [{\citenamefont {Monroe}\ \emph {et~al.}(2021)\citenamefont {Monroe}, \citenamefont {Campbell}, \citenamefont {Duan}, \citenamefont {Gong}, \citenamefont {Gorshkov}, \citenamefont {Hess}, \citenamefont {Islam}, \citenamefont {Kim}, \citenamefont {Linke}, \citenamefont {Pagano}, \citenamefont {Richerme}, \citenamefont {Senko},\ and\ \citenamefont {Yao}}]{Monore2021Programmable}%
  \BibitemOpen
  \bibfield  {author} {\bibinfo {author} {\bibfnamefont {C.}~\bibnamefont {Monroe}}, \bibinfo {author} {\bibfnamefont {W.~C.}\ \bibnamefont {Campbell}}, \bibinfo {author} {\bibfnamefont {L.-M.}\ \bibnamefont {Duan}}, \bibinfo {author} {\bibfnamefont {Z.-X.}\ \bibnamefont {Gong}}, \bibinfo {author} {\bibfnamefont {A.~V.}\ \bibnamefont {Gorshkov}}, \bibinfo {author} {\bibfnamefont {P.~W.}\ \bibnamefont {Hess}}, \bibinfo {author} {\bibfnamefont {R.}~\bibnamefont {Islam}}, \bibinfo {author} {\bibfnamefont {K.}~\bibnamefont {Kim}}, \bibinfo {author} {\bibfnamefont {N.~M.}\ \bibnamefont {Linke}}, \bibinfo {author} {\bibfnamefont {G.}~\bibnamefont {Pagano}}, \bibinfo {author} {\bibfnamefont {P.}~\bibnamefont {Richerme}}, \bibinfo {author} {\bibfnamefont {C.}~\bibnamefont {Senko}},\ and\ \bibinfo {author} {\bibfnamefont {N.~Y.}\ \bibnamefont {Yao}},\ }\bibfield  {title} {\bibinfo {title} {Programmable quantum simulations of spin systems with trapped ions},\ }\href {https://doi.org/10.1103/RevModPhys.93.025001}
  {\bibfield  {journal} {\bibinfo  {journal} {Rev. Mod. Phys.}\ }\textbf {\bibinfo {volume} {93}},\ \bibinfo {pages} {025001} (\bibinfo {year} {2021})}\BibitemShut {NoStop}%
\bibitem [{\citenamefont {Georgescu}(2020)}]{Georgescu2020Trapped}%
  \BibitemOpen
  \bibfield  {author} {\bibinfo {author} {\bibfnamefont {I.}~\bibnamefont {Georgescu}},\ }\bibfield  {title} {\bibinfo {title} {Trapped ion quantum computing turns 25},\ }\href {https://doi.org/10.1038/s42254-020-0189-1} {\bibfield  {journal} {\bibinfo  {journal} {Nat. Rev. Phys.}\ }\textbf {\bibinfo {volume} {2}},\ \bibinfo {pages} {278} (\bibinfo {year} {2020})}\BibitemShut {NoStop}%
\bibitem [{\citenamefont {Kjaergaard}\ \emph {et~al.}(2020)\citenamefont {Kjaergaard}, \citenamefont {Schwartz}, \citenamefont {Braum{\"u}ller}, \citenamefont {Krantz}, \citenamefont {Wang}, \citenamefont {Gustavsson},\ and\ \citenamefont {Oliver}}]{Kjaergaard2020Superconducting}%
  \BibitemOpen
  \bibfield  {author} {\bibinfo {author} {\bibfnamefont {M.}~\bibnamefont {Kjaergaard}}, \bibinfo {author} {\bibfnamefont {M.~E.}\ \bibnamefont {Schwartz}}, \bibinfo {author} {\bibfnamefont {J.}~\bibnamefont {Braum{\"u}ller}}, \bibinfo {author} {\bibfnamefont {P.}~\bibnamefont {Krantz}}, \bibinfo {author} {\bibfnamefont {J.~I.-J.}\ \bibnamefont {Wang}}, \bibinfo {author} {\bibfnamefont {S.}~\bibnamefont {Gustavsson}},\ and\ \bibinfo {author} {\bibfnamefont {W.~D.}\ \bibnamefont {Oliver}},\ }\bibfield  {title} {\bibinfo {title} {Superconducting qubits: Current state of play},\ }\href {https://doi.org/10.1146/annurev-conmatphys-031119-050605} {\bibfield  {journal} {\bibinfo  {journal} {Annu. Rev. Condens. Matter Phys.}\ }\textbf {\bibinfo {volume} {11}},\ \bibinfo {pages} {369} (\bibinfo {year} {2020})}\BibitemShut {NoStop}%
\bibitem [{\citenamefont {Arute}\ \emph {et~al.}(2019)\citenamefont {Arute}, \citenamefont {Arya}, \citenamefont {Babbush}, \citenamefont {Bacon}, \citenamefont {Bardin}, \citenamefont {Barends}, \citenamefont {Biswas}, \citenamefont {Boixo}, \citenamefont {Brandao}, \citenamefont {Buell} \emph {et~al.}}]{Arute2019Quantum}%
  \BibitemOpen
  \bibfield  {author} {\bibinfo {author} {\bibfnamefont {F.}~\bibnamefont {Arute}}, \bibinfo {author} {\bibfnamefont {K.}~\bibnamefont {Arya}}, \bibinfo {author} {\bibfnamefont {R.}~\bibnamefont {Babbush}}, \bibinfo {author} {\bibfnamefont {D.}~\bibnamefont {Bacon}}, \bibinfo {author} {\bibfnamefont {J.~C.}\ \bibnamefont {Bardin}}, \bibinfo {author} {\bibfnamefont {R.}~\bibnamefont {Barends}}, \bibinfo {author} {\bibfnamefont {R.}~\bibnamefont {Biswas}}, \bibinfo {author} {\bibfnamefont {S.}~\bibnamefont {Boixo}}, \bibinfo {author} {\bibfnamefont {F.~G.}\ \bibnamefont {Brandao}}, \bibinfo {author} {\bibfnamefont {D.~A.}\ \bibnamefont {Buell}}, \emph {et~al.},\ }\bibfield  {title} {\bibinfo {title} {Quantum supremacy using a programmable superconducting processor},\ }\href {https://doi.org/10.1038/s41586-019-1666-5} {\bibfield  {journal} {\bibinfo  {journal} {Nature}\ }\textbf {\bibinfo {volume} {574}},\ \bibinfo {pages} {505} (\bibinfo {year} {2019})}\BibitemShut {NoStop}%
\bibitem [{\citenamefont {Ren}\ \emph {et~al.}(2022)\citenamefont {Ren}, \citenamefont {Li}, \citenamefont {Xu}, \citenamefont {Wang}, \citenamefont {Jiang}, \citenamefont {Jin}, \citenamefont {Zhu}, \citenamefont {Chen}, \citenamefont {Song}, \citenamefont {Zhang} \emph {et~al.}}]{Ren2022Experimental}%
  \BibitemOpen
  \bibfield  {author} {\bibinfo {author} {\bibfnamefont {W.}~\bibnamefont {Ren}}, \bibinfo {author} {\bibfnamefont {W.}~\bibnamefont {Li}}, \bibinfo {author} {\bibfnamefont {S.}~\bibnamefont {Xu}}, \bibinfo {author} {\bibfnamefont {K.}~\bibnamefont {Wang}}, \bibinfo {author} {\bibfnamefont {W.}~\bibnamefont {Jiang}}, \bibinfo {author} {\bibfnamefont {F.}~\bibnamefont {Jin}}, \bibinfo {author} {\bibfnamefont {X.}~\bibnamefont {Zhu}}, \bibinfo {author} {\bibfnamefont {J.}~\bibnamefont {Chen}}, \bibinfo {author} {\bibfnamefont {Z.}~\bibnamefont {Song}}, \bibinfo {author} {\bibfnamefont {P.}~\bibnamefont {Zhang}}, \emph {et~al.},\ }\bibfield  {title} {\bibinfo {title} {Experimental quantum adversarial learning with programmable superconducting qubits},\ }\href {https://doi.org/10.1038/s43588-022-00351-9} {\bibfield  {journal} {\bibinfo  {journal} {Nat. Comput. Sci.}\ }\textbf {\bibinfo {volume} {2}},\ \bibinfo {pages} {711} (\bibinfo {year} {2022})}\BibitemShut {NoStop}%
\bibitem [{\citenamefont {Saffman}\ \emph {et~al.}(2010)\citenamefont {Saffman}, \citenamefont {Walker},\ and\ \citenamefont {M\o{}lmer}}]{Saffman2010Quantum}%
  \BibitemOpen
  \bibfield  {author} {\bibinfo {author} {\bibfnamefont {M.}~\bibnamefont {Saffman}}, \bibinfo {author} {\bibfnamefont {T.~G.}\ \bibnamefont {Walker}},\ and\ \bibinfo {author} {\bibfnamefont {K.}~\bibnamefont {M\o{}lmer}},\ }\bibfield  {title} {\bibinfo {title} {Quantum information with rydberg atoms},\ }\href {https://doi.org/10.1103/RevModPhys.82.2313} {\bibfield  {journal} {\bibinfo  {journal} {Rev. Mod. Phys.}\ }\textbf {\bibinfo {volume} {82}},\ \bibinfo {pages} {2313} (\bibinfo {year} {2010})}\BibitemShut {NoStop}%
\bibitem [{\citenamefont {Bluvstein}\ \emph {et~al.}(2024)\citenamefont {Bluvstein}, \citenamefont {Evered}, \citenamefont {Geim}, \citenamefont {Li}, \citenamefont {Zhou}, \citenamefont {Manovitz}, \citenamefont {Ebadi}, \citenamefont {Cain}, \citenamefont {Kalinowski}, \citenamefont {Hangleiter} \emph {et~al.}}]{Bluvstein2024Logical}%
  \BibitemOpen
  \bibfield  {author} {\bibinfo {author} {\bibfnamefont {D.}~\bibnamefont {Bluvstein}}, \bibinfo {author} {\bibfnamefont {S.~J.}\ \bibnamefont {Evered}}, \bibinfo {author} {\bibfnamefont {A.~A.}\ \bibnamefont {Geim}}, \bibinfo {author} {\bibfnamefont {S.~H.}\ \bibnamefont {Li}}, \bibinfo {author} {\bibfnamefont {H.}~\bibnamefont {Zhou}}, \bibinfo {author} {\bibfnamefont {T.}~\bibnamefont {Manovitz}}, \bibinfo {author} {\bibfnamefont {S.}~\bibnamefont {Ebadi}}, \bibinfo {author} {\bibfnamefont {M.}~\bibnamefont {Cain}}, \bibinfo {author} {\bibfnamefont {M.}~\bibnamefont {Kalinowski}}, \bibinfo {author} {\bibfnamefont {D.}~\bibnamefont {Hangleiter}}, \emph {et~al.},\ }\bibfield  {title} {\bibinfo {title} {Logical quantum processor based on reconfigurable atom arrays},\ }\href {https://doi.org/10.1038/s41586-023-06927-3} {\bibfield  {journal} {\bibinfo  {journal} {Nature}\ }\textbf {\bibinfo {volume} {626}},\ \bibinfo {pages} {58} (\bibinfo {year} {2024})}\BibitemShut {NoStop}%
\end{thebibliography}%


\begin{thebibliography}{1}%
\makeatletter
\providecommand \@ifxundefined [1]{%
 \@ifx{#1\undefined}
}%
\providecommand \@ifnum [1]{%
 \ifnum #1\expandafter \@firstoftwo
 \else \expandafter \@secondoftwo
 \fi
}%
\providecommand \@ifx [1]{%
 \ifx #1\expandafter \@firstoftwo
 \else \expandafter \@secondoftwo
 \fi
}%
\providecommand \natexlab [1]{#1}%
\providecommand \enquote  [1]{``#1''}%
\providecommand \bibnamefont  [1]{#1}%
\providecommand \bibfnamefont [1]{#1}%
\providecommand \citenamefont [1]{#1}%
\providecommand \href@noop [0]{\@secondoftwo}%
\providecommand \href [0]{\begingroup \@sanitize@url \@href}%
\providecommand \@href[1]{\@@startlink{#1}\@@href}%
\providecommand \@@href[1]{\endgroup#1\@@endlink}%
\providecommand \@sanitize@url [0]{\catcode `\\12\catcode `\$12\catcode `\&12\catcode `\#12\catcode `\^12\catcode `\_12\catcode `\%12\relax}%
\providecommand \@@startlink[1]{}%
\providecommand \@@endlink[0]{}%
\providecommand \url  [0]{\begingroup\@sanitize@url \@url }%
\providecommand \@url [1]{\endgroup\@href {#1}{\urlprefix }}%
\providecommand \urlprefix  [0]{URL }%
\providecommand \Eprint [0]{\href }%
\providecommand \doibase [0]{https://doi.org/}%
\providecommand \selectlanguage [0]{\@gobble}%
\providecommand \bibinfo  [0]{\@secondoftwo}%
\providecommand \bibfield  [0]{\@secondoftwo}%
\providecommand \translation [1]{[#1]}%
\providecommand \BibitemOpen [0]{}%
\providecommand \bibitemStop [0]{}%
\providecommand \bibitemNoStop [0]{.\EOS\space}%
\providecommand \EOS [0]{\spacefactor3000\relax}%
\providecommand \BibitemShut  [1]{\csname bibitem#1\endcsname}%
\let\auto@bib@innerbib\@empty
%</preamble>
\bibitem [{\citenamefont {Caro}\ \emph {et~al.}(2023)\citenamefont {Caro}, \citenamefont {Hinsche}, \citenamefont {Ioannou}, \citenamefont {Nietner},\ and\ \citenamefont {Sweke}}]{Caro2023Classical}%
  \BibitemOpen
  \bibfield  {author} {\bibinfo {author} {\bibfnamefont {M.~C.}\ \bibnamefont {Caro}}, \bibinfo {author} {\bibfnamefont {M.}~\bibnamefont {Hinsche}}, \bibinfo {author} {\bibfnamefont {M.}~\bibnamefont {Ioannou}}, \bibinfo {author} {\bibfnamefont {A.}~\bibnamefont {Nietner}},\ and\ \bibinfo {author} {\bibfnamefont {R.}~\bibnamefont {Sweke}},\ }\href {https://doi.org/10.48550/arXiv.2306.04843} {\bibinfo {title} {Classical {{Verification}} of {{Quantum Learning}}}} (\bibinfo {year} {2023})\BibitemShut {NoStop}%
\end{thebibliography}%

\end{document}

\title{Supplementary Materials for: Classical Verification of Quantum Learning Advantages with Noises}

\author{Yinghao Ma}
 \affiliation{Center for Quantum Information, IIIS, Tsinghua University, Beijing 100084, China}
 
\author{Jiaxi Su}
 \affiliation{Center for Quantum Information, IIIS, Tsinghua University, Beijing 100084, China}

\author{Dong-Ling Deng}
\email{dldeng@tsinghua.edu.cn}
\affiliation{Center for Quantum Information, IIIS, Tsinghua University, Beijing 100084,  China}
\affiliation{Hefei National Laboratory, Hefei, China}
\affiliation{Shanghai Qi Zhi Institute, No. 701 Yunjin Road, Xuhui District, Shanghai 200232, China}

\maketitle

In this Supplementary Materials, we provide technical details for the proofs of the theorems in the main text. For completeness and convenience, we introduce some lemmas that have been proved in the literature and restate the theorems in the main text.

\section{Lemma}
We present two mathematical lemmas that assist in the proofs of subsequent theorems. The first lemma, cited from Ref\cite{Caro2023Classical}, establishes that we can effectively approximate a probability distribution in $\ell_\infty$ distance using the empirical distribution of samples from it. The second lemma demonstrates how a random example oracle can be utilized to determine the Fourier coefficients of specified items.

\begin{lemma}[Lemma 3 in Ref. \cite{Caro2023Classical}]
    \label{SL sample}
    Let $q:\{0,1\}^n \rightarrow [0,1]$ be a probability distribution and let $(\eps,\delta) \in (0,1)^2$. There exists an efficient classical algorithm that, given $k=O\left(\frac{\log(1/\delta)}{\eps^2}\right)$ i.i.d. samples from $q$, it outputs the succinctly represented empirical distribution $\tilde q$ which is defined as:
    \begin{equation}
        \tilde q(s)=\frac{1}{k}\sum_{t\in T}\I(s=t)\ ,\,T\text{ denotes for the list of samples},
    \end{equation}
    such that $\Norm[\infty]{q-\tilde q}\le \eps$ with probability at least $1-\delta$. The algorithm's time and space complexities are both $\tilde O\left(\frac{1\log(1/\delta)}{\eps^2}\right)$.
\end{lemma}

\begin{lemma}
    \label{SL hat g}
    Let $f:\mc X_n\to\{0,1\}$ be a boolean function and let $(\eps,\delta)\in(0,1)^2$. Let $\hat g$ be the Fourier coefficient of $g=1-2f$. There exists an efficient classical algorithm that given a set $S\subseteq\mc X_n$, it takes $k=O\left(\frac{\log(|S|/\delta)}{\eps^2}\right)$ classical random examples to produce an estimation $\tilde{g}(s)$ of $\hat g(s)$ for $s\in S$, such that the probability of $|\hat g(s)-\tilde{g}(s)|\le \eps$ for all $s\in S$ is at least $1-\delta$. The algorithm's time and space complexities are $\tilde O\left(\frac{n|S|\log(|S|/\delta)}{\eps^2}\right)$ and $\tilde O\left(\frac{n\log(|S|/\delta)}{\eps^2}+n|S|\right)$ respectively.
\end{lemma}

\begin{proof}
    Let $T$ be the list of classical random examples, define:
    \begin{equation}
        \tilde{g}(s)=\frac{1}{k}\sum_{x\in T}g(x)\chi_s(x) , s\in L.
        \label{SEqu 1}
    \end{equation}
    By Hoeffding inequalities, for any $s\in S$:
    \begin{equation}
        \Pr(|\hat g(s)-\tilde{g}(s)|>\eps)\le2\e^{-\frac{1}{2}k\eps^2}.
    \end{equation}
    Let $2\e^{-\frac{1}{2}k\eps^2}\le\frac{\delta}{|S|}$, so $k=O\left(\frac{\log(|S|/\delta)}{\eps^2}\right)$, then the probability of success is at least $\Pr(\forall s\in S,|\hat g(s)-\tilde{g}(s)|\le\eps)\ge 1-\frac{\delta}{|S|}\cdot|S|=1-\delta$.

    \textit{Time complexity.}
    For each $s\in S$, we use equation \ref{SEqu 1} to compute $\tilde{g}(s)$. The summation over $T$ in \ref{SEqu 1} takes $\tilde O(kn)$ time since we need $O(n)$ time to compute each $\chi_s(x)$ and $\tilde O(k)$ time for summation, and the final division only takes $\tilde O(1)$ time. So the time complexity of this algorithm is $|S|\cdot\tilde O(kn)=\tilde O\left(\frac{n|S|\log(|S|/\delta)}{\eps^2}\right)$.
    
    \textit{Space complexity.} The algorithm needs $O(|S|n+kn)$ space for input data, $\tilde O(1)+O(n)$ space for computation and $\tilde O(|S|)$ for output. So the space complexity of this algorithm is $\tilde O(|S|n+kn)=\tilde O\left(\frac{n\log(|S|/\delta)}{\eps^2}+n|S|\right)$.
\end{proof}

\section{Proof of Theorem 1}

We now present the formal statement of Theorem 1 along with its proof. To make the proof more concise, we extract one step of the mathematical calculations in the proof and present it as a separate lemma.

\begin{theorem}[Theorem 1 in the main text]
    \label{ST noise1}
    Let $p_0:\{0,1\}^n \rightarrow [0,1]$ be a probability distribution over $\mc X_n=\{0,1\}^n$ and let $(\theta,\delta,\eta)\in(0,1)^2\times(0,\frac{1}{2})$. We define the noisy distribution as 
    \begin{equation}
        p_{\eta}(s)=\sum_{s'\in\mc X_n}\eta^{d_H(s,s')}(1-\eta)^{n-d_H(s,s')}p_0(s'),
    \end{equation}
    where $d_H(s,s')=\Norm[1]{s-s'}$ denotes for the Hamming distance between $s$ and $s'$.
    For $\eta\le \frac{1}{10}\theta$, there exists an efficient classical algorithm that takes $k=O\left(\frac{\log(n/\delta)}{\theta^2}\right)$ noisy samples from $p_{\eta}(s)$ and finds all possibly large components in $p_0(s)$ with success probability at least $1-\delta$. In particular, such an algorithm produces a set $L$ with $|L|\le\lfloor\frac{2}{\theta}\rfloor$ satisfying $H=\{s\in\mc X_n|p_0(s)\ge \theta\}\subseteq L$ with success probability no less than $1-\delta$. 
    The algorithm's time and space complexities are $\tilde O\left(\frac{n^2\log(n/\delta)}{\theta^3}\right)$ and $O\left(\frac{n\log(n/\delta)}{\theta^2}\right)$, respectively.
\end{theorem}

\begin{proof}
    We provide here the pseudo-code representation of the algorithm presented in Theorem 1.
    \begin{algorithm}[H]
        \caption{Classical Error-Rectification Algorithm}
        \begin{algorithmic}[1]
    	\Statex \textbf{input:} $T$ (the list of noisy samples), $n$, $\theta$
    	\Statex \textbf{output:} $L$
    	\Procedure{Classical Error-Rectification}{$T,n,\theta$}
    	\State $L_0\gets \{e\}$ where $e$ refers to an empty string of $\{0,1\}^*$
    	\For{$m \gets 1 \text{ to } n$}
    	\State $L_m'\gets L_{m-1}\times\{0,1\}$ (For each element in $L_m$, concatenate either $0$ or $1$ at the end.)
    	\For{$s_m$ in $L_m'$}
    	\State $s_m.freq\gets 0$
    	\EndFor 
    	\For{$t$ in $T$}
    	\State $t_m\gets$ the first $m$ bits of $t$
    	\State $s_m\gets$ the nearest one to $t_m$ in $L_m'$ (if there are multiple, choose one randomly)
            \State $s_m.freq\gets s_m.freq+\frac{1}{k}$
            \EndFor 
            \State Sort $L_m'$ in descending order with respect to $s_m.freq$
            \If{$|L_m'|\le \lfloor\frac{2}{\theta}\rfloor$}
            \State $L_m\gets L_m'$
            \Else 
            \State $L_m\gets$ first $\lfloor\frac{2}{\theta}\rfloor$ terms in $L_m'$
            \EndIf
            \EndFor
            \State $L\gets L_n$
            \State\Return $L$
            \EndProcedure
        \end{algorithmic}
        \label{SA 1}
    \end{algorithm}

    Denote $p_m(s_m)$ to be the original probability distribution for the first $m$ bits. Define $H_m=\{s_m\in\{0,1\}^m|p_m(s_m)\ge \theta\},H=H_n$. We will prove $H_m\subseteq L_m$ using mathematical induction on $m$. Initially $H_0=L_0=\{e\}$. If $H_{m-1}\subseteq L_{m-1}$ at step $m$, we have $H_m\subseteq L_m'$ because for each $s_m\in H_m$, the first $m-1$ bits $s_{m-1}$ satisfies $p_{m-1}(s_{m-1})\ge p_m(s_m)\ge \theta$, so $s_{m-1}\in H_{m-1}\subseteq L_{m-1}$, therefore $s_m\in L'_m$.
	
    For the matching progress in line 8-12 of the algorithm, consider the first $m$ bits $t_m$ of a noisy sample with the corresponding original noiseless sample being $s_m\in L'_m$. Assume the algorithm mismatches it with another $s_m'\ne s_m\in L_m'$, the noise must flip at least half of the different bits between $s_m$ and $s'_m$ to make the algorithm mismatch, so the probability is:
    \begin{equation}
        \Pr(s_m\to s'_m)\le\sum_{i=0}^{\lfloor d/2\rfloor}\left(1-\frac{1}{2}\I(i=d/2)\right){d\choose i}\eta^{d-i}(1-\eta)^{i}=P_d(\eta)\le \eta.
    \end{equation} 
    Here $d$ is the distance between $s$ and $s'$, and the factor $(1-\frac{1}{2}\I(i=d/2))$ means when the distances to $s$ and $s'$ are the same, the probability needs to be divided by at least $2$ because the algorithm randomly selects one if there are multiple nearest terms. The proof of $P_d(\eta)\le p$ is given in lemma \ref{SL P_d}. Then we have the probability of a correct match is at least:
    \begin{equation}
        \Pr(s_m\to s_m)\ge 1-|L'|\eta\ge 1-\frac{4\eta}{\theta}\ge\frac{3}{5}.
    \end{equation}

    Denote $p_m^{match}(s_m)$ to be the probability distribution of the matching result of a noisy sample, then for $s_m\in H_{m}$, we have $p_m^{match}(s_m)\ge\Pr(s_m\to s_m)p_m(s_m)\ge\frac{3}{5}\theta$. Consider the empirical distribution $p_m^{emp}(s_m)$ we get in line 11 of the algorithm, by lemma \ref{SL sample}, with probability at least $1-\frac{\delta}{n}$, $|p_m^{emp}(s_m)-p_m^{match}(s_m)|\le\frac{\theta}{10}$ for all $s_m\in L_m'$ using $k=O\left(\frac{\log(n/\delta)}{\theta^2}\right)$ samples. Combine these conclutions together we infer that with probability at least $1-\frac{\delta}{n}$, $p_m^{emp}(s_m)\ge\frac{\theta}{2}$ for all $s_m\in H_m$. Notice that there are at most $\left\lfloor\frac{2}{\theta}\right\rfloor$ of terms with $p_m^{emp}(s_m)\ge\frac{\theta}{2}$ in $L_m'$, this leads to the conclusion that $H_m\subseteq L_m$. Finally we have $H\subseteq L$ with probability at least:
    \begin{equation}
        \Pr(H\subseteq L)\ge 1- n\cdot \frac{\delta}{n}\ge 1-\delta.
    \end{equation}
    Another condition $|L|\le\lfloor\frac{2}{\theta}\rfloor$ can be directly inferred from the line 14-17 of the algorithm.

    \textit{Time complexity.} 
    For each iteration, line 8-12 takes $k\cdot\tilde O\left(n\cdot\frac{1}{\theta}\right)=\tilde O\left(\frac{n\log(n/\delta)}{\theta^3}\right)$ time (the notation $\tilde O$ hides the $\log k$ factor for storage and manipulation of $p_m.freq$), and the other parts needs $\tilde O\left(\frac{1}{\theta}+\frac{1}{\theta}\log \frac{1}{\theta}\right)=\tilde O\left(\frac{1}{\theta}\log \frac{1}{\theta}\right)$ time. So the space complexity of this algorithm is $n\cdot\left(\tilde O\left(\frac{n\log(n/\delta)}{\theta^3}\right)+\tilde O\left(\frac{1}{\theta}\log \frac{1}{\theta}\right)\right)+O\left(\frac{1}{\theta}\right)=\tilde O\left(\frac{n^2\log(n/\delta)}{\theta^3}\right)$.

    \textit{Space complexity.}
    The algorithm needs $kn=O\left(\frac{n\log(n/\delta)}{\theta^2}\right)$ space for $T$, $O\left(\frac{n}{\theta}\right)+\tilde O\left(\frac{1}{\theta}\right)$ space for $L_m,L_m'$ and $s_m.freq$, and $O(n)+\tilde O(1)$ space for $s,t$ and iteration counter. So the space complexity of this algorithm is $O\left(\frac{n\log(n/\delta)}{\theta^2}\right)+O\left(\frac{n}{\theta}\right)+\tilde O\left(\frac{1}{\theta}\right)+O(n)+\tilde O(1)=O\left(\frac{n\log(n/\delta)}{\theta^2}\right)$.
\end{proof}

\begin{lemma}
    \label{SL P_d}
    For $\eta \in(0,\frac{1}{2})$ and $d\in\mathbb{N}^*$, the polynomial
    \begin{equation}
        P_d(\eta)=\sum_{i=0}^{\lfloor d/2\rfloor}\left(1-\frac{1}{2}\I(i=d/2)\right){d\choose i}\eta^{d-i}(1-\eta)^{i}
    \end{equation}
    satisfies that $P_d(\eta)\le \eta$.
\end{lemma}

\begin{proof}
    For even number:
    \begin{align}
        P_{2d}&=\sum_{i=0}^{d-1}{2d\choose i}\eta^{2d-i}(1-\eta)^{i}+\frac{1}{2}{2d\choose d}\eta^{d}(1-\eta)^{d}\\
        &=\sum_{i=0}^{d-1}{2d-1\choose i}\eta^{2d-i}(1-\eta)^{i}+\sum_{i=1}^{d-1}{2d-1\choose i-1}\eta^{2d-i}(1-\eta)^{i}+{2d-1\choose d-1}\eta^d(1-\eta)^d\\
        &=\eta\sum_{i=0}^{d-1}{2d-1\choose i}\eta^{2d-i-1}(1-\eta)^{i}+(1-\eta)\sum_{i=0}^{d-2}{2d-1\choose i}\eta^{2d-i-1}(1-\eta)^{i}+{2d-1\choose d-1}\eta^d(1-\eta)^d\\
        &=\sum_{i=0}^{d-1}{2d-1\choose i}\eta^{2d-i-1}(1-\eta)^{i}=P_{2d-1}.
    \end{align}
    For odd number:
    \begin{align}
        P_{2d-1}&=\sum_{i=0}^{d-1}{2d-1\choose i}\eta^{2d-i-1}(1-\eta)^{i}
        \le\eta\sum_{i=0}^{d-1}{2d-1\choose i}\cdot\left(\frac{1}{2}\right)^{2d-2}\\
        &=\frac{1}{2}\eta\sum_{i=0}^{2d-1}{2d-1\choose i}\cdot\left(\frac{1}{2}\right)^{2d-2}
        =\frac{1}{2}\eta\cdot2^{2d-1}\cdot\left(\frac{1}{2}\right)^{2d-2}=\eta.
    \end{align}
    So for all $d\in\mathbb{N}^*$, $P_d(\eta)\le\eta$.
\end{proof}

Additionally,  we provide an improved version of Theorem \ref{ST noise1} that relax the restriction on noises. Theorem \ref{ST noise1} assumes independent bit flip noise, but this condition can be relaxed to a more general setting. Specifically, we consider a noise model described by a $2^n\times 2^n$ transition matrix $M_{ts}=\Pr(\text{noisy sample}=t|\text{original sample}=s)$, such that the noisy distribution $p_\eta=Mp_0$. We require the transition matrix $M$ to satisfy the following property: for any original sample $s$ and any $i\in [1,n]$, the probability of flipping the $i$-th bit of $s$ is at most $\eta$. This noise model allows for correlations between different bits and also permits the noise to depend on the original sample $s$. Theorem \ref{ST improved} presents the performance of Algorithm \ref{SA 1} under this noise model. The result indicate that with only a slight increase in the restriction on $\eta $ ($\eta\le\frac{1}{20}\theta$ instead of $\eta\le\frac{1}{10}\theta$), the algorithm achieves the same performance as stated in Theorem \ref{ST noise1}.

\begin{theorem}
    \label{ST improved}
    Let $p_0:\{0,1\}^n \rightarrow [0,1]$ be a probability distribution over $\mc X_n=\{0,1\}^n$ and let $(\theta,\delta,\eta)\in(0,1)^2\times(0,\frac{1}{2})$.
    Suppose that the samples from $p_0$ are subjected to a noise process with the following property: for any $i\in[n]$, the $i$-th bit of the sample is flipped with probability at most $\eta$.
    For $\eta\le \frac{1}{20}\theta$, there exists an efficient classical algorithm that takes $k=O\left(\frac{\log(n/\delta)}{\theta^2}\right)$ noisy samples and finds all possible large components in $p_0(s)$ with success probability at least $1-\delta$. In particular, such an algorithm produces a set $L$ with $|L|\le\lfloor\frac{2}{\theta}\rfloor$ satisfying $H=\{s\in\mc X_n|p_0(s)\ge \theta\}\subseteq L$ with success probability no less than $1-\delta$. 
    The algorithm's time and space complexities are $\tilde O\left(\frac{n^2\log(n/\delta)}{\theta^3}\right)$ and $O\left(\frac{n\log(n/\delta)}{\theta^2}\right)$, respectively. 
\end{theorem}

\begin{proof}
    We use the same algorithm as defined in Theorem \ref{ST noise1}. The proof is similar to the proof of Theorem \ref{ST noise1}, but here we need a different approach to upper bound the probability of mismatching, i.e. $\Pr(s_m\to s_m')$ for $s_m'\ne s_m$. Still we consider the algorithm mismatches $s_m\in L_m'$ with another $s_m'\in L_m'$, which means that the noise must flip at least half of the different bits between $s_m$ and $s'_m$. Denote $d$ to be the number of those different bits, and $x_i=\I(\text{the $i$-th of the different bits is flipped})$ for $i\in[1,d]$, since each bit is flipped with probability at most $\eta$, the probability of the mismatch:
    \begin{equation}
        \Pr(s_m\to s_m')\le\Pr(\text{at least half of the different bits are flipped})
        \le\E\left[\frac{2}{d}\sum_{i=1}^dx_i\right]
        =\frac{2}{d}\sum_{i=1}^d\E[x_i]
        \le\frac{2}{d}\cdot d\eta\le 2\eta.
    \end{equation}
    Then we have the probability of a correct match is at least:
    \begin{equation}
        \Pr(s_m\to s_m)\ge 1-|L'|2\eta\ge 1-\frac{8\eta}{\theta}\ge\frac{3}{5}.
    \end{equation}
    The remainder of the proof follows the same steps as in Theorem \ref{ST noise1}. Thus, we can obtain the same result as in Theorem \ref{ST noise1}.
\end{proof}

\section{Proof of Corollary 1}

With the result $L$ from Algorithm \ref{SA 1}, we can use the method in Lemma \ref{SL hat g} to estimate the Fourier coefficients of the elements in $L$, thereby obtaining an $\eps$-approximation of $\hat g(s)$ with respect to $\ell_\infty$ distance. The detailed proof of this result is provided in Theorem \ref{ST noise2}:

\begin{theorem}[Corollary 1 in the main text]
    \label{ST noise2}
    Let $f:\mc X_n\to\{0,1\}$ be a boolean function and let $(\eps,\delta)\in(0,1)^2$. Let $\hat g$ be the Fourier coefficient of $g=1-2f$. There exists an efficient classical algorithm that takes $k=O\left(\frac{\log(n/\delta)}{\eps^4}\right)$ samples from a noisy QFS circuit with noise strength $\eta\le\frac{1}{10}\eps^2$ and $k'=O\left(\frac{\log(1/\eps\delta)}{\eps^2}\right)$ classical random examples to produce a succinctly represented estimation $\tilde{g}$ of $\hat g$, such that $\Norm[0]{\tilde{g}}\le\frac{2}{\eps^2}$ and $\Norm[\infty]{\hat g-\tilde{g}}\le \eps$ with success probability at least $1-\delta$. 
    The algorithm's time and space complexities are $\tilde O\left(\frac{n^2\log(n/\delta)}{\eps^6}\right)$ and $O\left(\frac{n\log(n/\delta)}{\eps^4}\right)$, respectively.
\end{theorem}
\begin{proof}
    Use the algorithm in Theorem \ref{ST noise1}(replace $\theta$ by $\eps^2$) to find $L$ satisfying $H=\{s\in\mc X_n|p_0(s)\ge \eps^2\}\subseteq L$ with success probability $1-\frac{\delta}{2}$. Then use the algorithm in Lemma \ref{SL hat g} to produce an estimation $\tilde{g}(s)$ of $\hat g(s)$ for $s\in L$ such that $\forall s\in S,|\hat g(s)-\tilde{g}(s)|\le \eps$ with success probability at least $1-\frac{\delta}{2}$. By Theorem \ref{ST noise1} and Lemma \ref{SL hat g}, these require $k=O\left(\frac{\log(n/\delta)}{\eps^4}\right)$ samples from a noisy QFS circuit and $k'=O\left(\frac{\log(1/\eps\delta)}{\eps^2}\right)$ classical random examples respectively.
    We define $\tilde g(s)=0$ for $s\notin L$, thus $\Norm[0]{\tilde{g}}\le|L|\le\frac{2}{\eps^2}$.
    To show $\Norm[\infty]{\hat g-\tilde{g}}\le \eps$, notice that for $s\notin L$, $|\hat g(s)-\tilde{g}(s)|=|\hat g(s)|\le\eps$, therefore with probability at least $1-\frac{\delta}{2}\times 2=1-\delta$, we have $\Norm[\infty]{\hat g-\tilde{g}}\le \eps$. The time/space complexity of this algorithm is simply the sum of the time/space complexities of the two algorithms from Theorem \ref{ST noise1} and Lemma \ref{SL hat g}.
\end{proof}

\section{Proof of Theorem 2}

For completeness, we restate the formal definition of agnostic learning used in this paper:

\begin{definition}[Functional Agnostic Proper Learning]
    A hypothesis class $\mc H\subseteq \mc X_n^{\{0,1\}}$ is efficiently $\alpha$-agnostic proper learnable with respect to a function class $F\subseteq \mc X_n^{\{0,1\}}$ under the uniform input distribution $\mc U_n$, if there exist an efficient algorithm $A$ satisfying: for arbitrary $f\in F$ and $(\eps,\delta)\in(0,1)^2$, algorithm $A$ with access to oracle $O$ returns a hypothesis $h_0\in \mc H$ satisfying
    \begin{equation}
        \Pr_{x\sim \mc V_n}(f(x)\ne h_0(x))\le\alpha\cdot\min_{h\in \mc H}\Pr_{x\sim \mc V_n}(f(x)\ne h(x))+\eps
    \end{equation}
    with probability at least $1-\delta$. The algorithm should be both efficient(i.e. $poly(n,1/{\eps},1/{\delta}$) in time and sample complexity.
\end{definition}

As presented in Theorem \ref{ST pac}, the parity class $H_p=\{h(x)=s\cdot x|s\in\mc X_n\}$ is efficiently $1$-agnostic proper learnable through the collaboration of a noisy quantum device and a classical computer, each having access to a quantum example oracle and random example oracle respectively:

\begin{theorem}[Theorem 2 in the main text]
    \label{ST pac}
    Let $f:\mc X_n\to\{0,1\}$ be a boolean function and let $(\eps,\delta)\in(0,1)^2$. There exists an efficient classical algorithm that takes $k=O\left(\frac{\log(n/\delta)}{\eps^4}\right)$ samples from noisy QFS circuit with error strength $\eta\le\frac{1}{10}\eps^2$ and $k'=O\left(\frac{\log(1/\eps\delta)}{\eps^2}\right)$ classical random examples to output $s_0\in\mc X_n$ such that
    \begin{equation}
        \Pr_{x\sim \mc V_n}(f(x)\ne s_0\cdot x)\le\min_{s\in\mc X_n}\Pr_{x\sim \mc V_n}(f(x)\ne s_0\cdot x)+\eps
    \end{equation}
    with success probability at least $1-\delta$. 
    The algorithm's time and space complexities are $\tilde O\left(\frac{n^2\log(n/\delta)}{\eps^6}\right)$ and $O\left(\frac{n\log(n/\delta)}{\eps^4}\right)$, respectively.
\end{theorem}

\begin{proof}
    As proved in Corollary \ref{ST noise2}, we can obtain a succinctly represented estimation $\tilde g$ of $\hat g$ with success probability at least $1-\delta$, such that $\Norm[\infty]{\hat g-\tilde{g}}\le \eps$ and $\Norm[0]{\tilde g}\le \frac{2}{\eps^2}$. The algorithm requires $k=O\left(\frac{\log(n/\delta)}{\eps^4}\right)$ samples from noisy QFS circuit with error strength $\eta\le\frac{1}{10}\eps^2$ and $k'=O\left(\frac{\log(1/\eps\delta)}{\eps^2}\right)$ classical random examples. Let $s_0\in\arg\max_s\tilde{g}(s)$ and $s_1\in\arg\max_s\hat{g}(s)$, then $\hat g(s_0)\ge\tilde{g}(s_0)-\eps\ge\tilde{g}(s_1)-\eps\ge\hat{g}(s_1)-2\eps=\max_s\hat g(s)-2\eps$, so $\hat g(s_0)$ is $2\eps$-approximation maximal.
    
    Notice that the loss function can be written as:
    \begin{align}
        L_f(s)&=\Pr_{x\sim \mc V_n}(s\cdot x\ne f(x))
        =\Pr_{x\sim \mc V_n}(\chi_s(x)\ne g(x))\\
        &=\frac{1}{2^n}\sum_{x\in\mc X_n}\frac{1-g(x)\chi_s(x)}{2}
        =\frac{1-\hat{g}(s)}{2},
    \end{align}
    therefore we have $\Pr_{x\sim \mc V_n}(f(x)\ne s_0\cdot x)\le\min_{s\in\mc X_n}\Pr_{x\sim \mc V_n}(f(x)\ne s_0\cdot x)+\eps$. The algorithm's time and space complexities are $\tilde O\left(\frac{n^2\log(n/\delta)}{\eps^6}\right)$ and $O\left(\frac{n\log(n/\delta)}{\eps^4}\right)$, respectively, same as the algorithm in Corollary \ref{ST noise2}.
\end{proof}

\section{Proof of Theorem 3}

For completeness, we present the formal definition of interactive proof system for agnostic learning used in this paper:

\begin{definition}[Interactive Proof System]
    A hypothesis class $\mc H$ is efficiently $\alpha$-agnostic proper verifiable with respect to function class $F$ under the uniform input distribution, if there exist a pair of efficient algorithms $(V,P)$ such that, given $(\eps,\delta)\in(0,1)^2$ and access for oracle $O_V,O_P$ respectively, it satisfies the following properties for arbitrary $f\in F$:
    \begin{itemize}
        \item \textbf{Completeness: } If $V$ interacts with the honest prover $P$, then with probability at least $1-\delta$, $V$ accepts the interaction and outputs a correct answer.
        \item \textbf{Soundness: } For arbitrary prover $P'$ that interacts with $V$, even with unlimited computational power and knowledge about the target function, with probability at most $\delta$, $V$ accepts the interaction and outputs a wrong answer.
    \end{itemize}
    Here a correct answer refers to a hypothesis $h_0\in \mc H$ satisfying
    \begin{equation}
        \Pr_{x\sim \mc V_n}(f(x)\ne h_0(x))\le\alpha\cdot\min_{h\in \mc H}\Pr_{x\sim \mc V_n}(f(x)\ne h(x))+\eps.
    \end{equation}
\end{definition}

Building on the learning algorithm from Theorem \ref{ST pac}, we present a classical-quantum interactive proof protocol for $1$-agnostic parity learning. The outcomes of our work are outlined in Theorem \ref{ST interact}:

\begin{definition}
    Let $\tau\in(0,1)$, we define $F_\tau=\{f\in\mc X_n^{\{0,1\}}:\forall s\in\mc X_n,|\hat g(s)|\ge\tau\lor\hat g(s)=0\}$ to be the class of boolean functions with no non-zero Fourier coefficients smaller than $\tau$.
\end{definition}

\begin{theorem}[Theorem 3 in the main text]
    \label{ST interact}
    Let $(\tau,\eps,\delta)\in(0,1)^3$. The parities of boolean functions in $F_\tau$ is 1-agnostic proper verifiable by a classical verifier $V$ and a noisy quantum prover $P$ with error strength $\eta\le \frac{1}{10}\tau^2$, each having access to a quantum example oracle and random example oracle respectively, using only 1 round of classical communication. 
\end{theorem}
\begin{proof}
    The interactive proof system $(V,P)$ is defined as follows: (assume samples in different steps are independent)
    \begin{enumerate}
        \item Based on the algorithm in Theorem \ref{ST noise1}, the verifier $V$ asks the prover $P$ to provide $k=O\left(\frac{\log(n/\delta)}{\tau^4}\right)$ samples from noisy QFS circuit, and produce a set $L$ containing $H=\{s\in\mc X_n:|\hat g(s)|\ge \tau\}=\{s\in\mc X_n:|\hat g(s)|\ne 0\}$ with probability at least $1-\frac{\delta}{3}$, using samples from $P$ and $k'_1=O\left(\frac{\log(1/\tau\delta)}{\tau^2}\right)$ samples from random example oracle.
        \item The verifier $V$ checks the reliability of $L$ by using the algorithm in Lemma \ref{SL hat g} to obtain a $\frac{1}{8}\tau^3$-approximation $\tilde{g}(s)$ of $\hat g(s)$ for $s\in L$ with probability $1-\frac{\delta}{3}$, which requires $k'_2=O\left(\frac{\log(1/\tau\delta)}{\tau^6}\right)$ samples from random example oracle. If the sum $\tilde S=\sum_{s\in L}(\tilde g(s))^2\ge 1-\frac{1}{2}\tau^2$, the verifier chooses to trust $H\subseteq L$, otherwise it rejects the interaction.
        \item The verifier $V$ again uses the algorithm in Lemma \ref{SL hat g} to obtain an $\eps$-approximation $\tilde{g}'(s)$ of $\hat g(s)$ for $s\in L$ with probability $1-\frac{\delta}{3}$ using $k'_3=O\left(\frac{\log(1/\tau\delta)}{\eps^2}\right)$ samples from random example oracle. Then $V$ chooses $s_0\in\arg\max_{s\in L}\tilde g(s)$ and outputs $h(x)=s_0\cdot x$ as learning result.
    \end{enumerate}
     
    For Step 2, let $S=\sum_{s\in L}(\hat g(s))^2$. Based on $|\hat g(s)-\tilde{g}(s)|\le\frac{1}{8}\tau^3$ for all $s\in L$ and $|L|\le\frac{2}{\tau^2}$, we have:
    \begin{align}
        |S-\tilde S|&\le\sum_{s\in L}|(\hat{g}(s))^2-(\tilde{g}(s))^2|
        =\sum_{s\in L}|(\hat{g}(s)-\tilde{g}(s))|\cdot|(\hat{g}(s)+\tilde{g}(s))|\\
        &\le\sum_{s\in L}\frac{1}{8}\tau^3\left(2|\hat{g}(s)|+\frac{1}{8}\tau^3\right)
        =\frac{1}{4}\tau^3\left(\sum_{s\in L}|\hat{g}(s)|\right)+\frac{1}{64}\tau^6\cdot|L|\\
        &\le\frac{1}{4}\tau^3\left(|L|\sum_{s\in L}(\hat{g}(s))^2\right)^{\frac{1}{2}}+\frac{1}{32}\tau^4
        \le\frac{1}{2\sqrt{2}}\tau^2+\frac{1}{32}\tau^4<\frac{1}{2}\tau^2
    \end{align}
    If $H\subseteq L$, then $S=1$, $\tilde S\ge1-\frac{1}{2}\tau^2$, otherwise $S\le1-\tau^2$, $\tilde S<1-\frac{1}{2}\tau^2$. So with probability at least $1-\frac{\delta}{3}$, $L$ passes the validation if and only if $H\subseteq L$.

    For Step 3, if $H\subseteq L$, with probability at least $1-\frac{\delta}{3}$ we will find an $\eps$-approximation $\tilde{g}'(s)$ of $\hat g(s)$ for $s\in L$, following the proof of Theorem \ref{ST pac} we can see that $V$ finds a correct answer for the learning task.

    \textit{Completeness.} 
    If $P$ is a honest quantum prover, then each step of the protocol successes with probability at least $1-\frac{\delta}{3}$, so
    \begin{equation}
        \Pr(V\text{ returns a correct answer when interacting with an honest }P)\ge1-\frac{\delta}{3}\cdot3=1-\delta
    \end{equation}

    \textit{Soundness.} 
    Assume $L$ is already decided in Step 1, we consider the probability space over the samples in Step 2 and 3: assume $V$ gives a wrong answer, if $H\subseteq L$, then it means Step 3 fails, otherwise $H\not\subseteq L$ but $L$ passes the validation in Step 2, both situation have probability $\le\frac{\delta}{3}$, so
    \begin{equation}
        \Pr(V\text{ returns a wrong answer when interacting with arbitrary }P')\le\frac{\delta}{3}\le\delta
    \end{equation}
\end{proof}

\bibliography{ref}